\documentclass[]{raa}
\usepackage{graphicx,times}
\usepackage{natbib}
\usepackage{amssymb,amsmath}
\usepackage{rotating}
\usepackage{longtable}
\usepackage{supertabular}
\usepackage{longtable,lscape}
\bibpunct{(}{)}{;}{a}{}{,}

\def\aap{A\&A}

\def\apjs{ApJS}
\def\apjl{ApJL}

\usepackage[a4paper=true,dvipdfm=true,pagebackref=true]{hyperref}
\hypersetup{pdftitle = The title of my PDF, pdfauthor = My name, pdfsubject= The subject, pdfkeywords = keyword1 keyword2 keyword3}
\hypersetup{colorlinks = true, linkcolor = green, anchorcolor = red, citecolor = blue, filecolor = red, pagecolor = red, urlcolor = red}

\begin{document}

   \title{Long-term optical-infrared color variability of  blazars
$^*$
\footnotetext{\small $*$ Supported by the National Natural Science Foundation of China.}
}

 \volnopage{ {\bf 2012} Vol.\ {\bf X} No. {\bf XX}, 000--000}
   \setcounter{page}{1}

   \author{Bing-Kai Zhang\inst{1}, Xiao-Shan Zhou\inst{1}, Xiao-Yun Zhao\inst{1}, Ben-Zhong Dai
      \inst{2}
   }

   \institute{ Department of Physics, Fuyang Normal College, Fuyang
              236041, China; {\it zhangbk@ihep.ac.cn}\\
        \and
              Department of Physics, Yunnan University, Kunming 650091,
             China\\
\vs \no
   {\small Received ; accepted }
}

\abstract{
The long-term optical and infrared color variability of blazars has been investigated with the SMARTS monitoring data.
 The sample in this study consists of 49 flat spectrum radio quasars and 22 BL Lac objects.
  The fractional variability amplitudes of each source have been calculated in both optical $R$ band and infrared $J$ band. Overall, the variability amplitudes of FSRQs are larger than those of BL Lac objects. The results also suggest that the variability amplitude of most FSRQs is larger at lower energy band ($J$ band) than at higher one ($R$ band), while  the variability amplitude of BL Lacs are larger at higher energy band.  Two types of blazars both display color variation along with  the variability in brightness. However, they  show different variation behaviors in general.  With the whole data set, 35 FSRQs exhibit redder-when-brighter trends, and 11 FSRQs  exhibit opposite trends; 11 BL Lacs follow bluer-when-brighter trends, and 7 BL Lacs follow opposite trends.
 The examination in detail shows that there are 10 blazars showing redder-when-brighter trend in their low state, and bluer-when-brighter or stable-when-brighter trends in their high state. Some more complicated color behaviors have also been detected in several blazars. The non-thermal jet emission and the thermal emission from the accretion disc are employed to explain the observed color behaviors.
\keywords{galaxies: active -- BL Lacertae objects: general -- quasars: general -- methods: statistical
}
}

   \authorrunning{B.-K. Zhang et al. }            
   \titlerunning{Long-term optical-infrared color variability of blazars}  
   \maketitle

%
\section{Introduction}           
\label{sect:intro}

Blazars consist of flat spectrum radio quasars (FSRQs) and BL Lacertae objects (BL Lacs).
They are dominated by non-thermal continuum emission.  The flux decreases along with the frequency increasing, which approximately follows a power law.
Measurements of the spectral energy distribution (SEDs) play an important role in understanding the emission processes of blazars.   Since blazars vary rapidly and violently at all wavelengths from radio to $\gamma$-ray with time scales from minutes to years \citep{carini92,arevalo08,ghosh00,ciprini07,chatterjee12,zhang13,dammando13}, the lack of simultaneous data at the various wavelengths is an obstacle in the broadband study of blazars.  Flux variations in blazars are usually accompanied by changes in the spectral shape.  Therefore, the analysis of spectral changes of blazars can put some strong constraints on the physical processes that are responsible for these variations.

The spectral change can be characterized by the color change.
Color variations in blazars have been explored by several authors.
 For 28 radio-selected BL Lacs in \cite{fan00}, two showed flatter spectra when getting brighter, while three showed opposite behavior.
\cite{vagne03} found bluer-when-brighter (BWB) trends in the $BVRI$ bands  for all 8 BL Lacs in their sample.
\cite{gu06} found that 3 out of 5 BL Lacs exhibited significant BWB trends and 2 out of 3 FSRQs exhibited redder-when-brighter (RWB) trends. However, \cite{gu11} showed that only one FSRQs is RWB in a sample of 29 SDSS FSRQs.
\cite{rani10} found that 3 out of 6 BL Lacs tended to become bluer with increase in brightness,  and 4 out of 6 FSRQs exhibited opposite trends.
\cite{ikeji11} found that 28 out of 32 well observed blazars showed BWB trend.
\cite{bonning12} found that FSRQs had redder optical-infrared colors when they were brighter, while BL Lac objects showed no such trend.
\cite{gaur12} suggested that 6 out of 10 BL Lacs followed the BWB trend, while all 6 BL Lacs in \cite{sandrinelli14} showed no BWB trends.
The color behaviors of some individual sources are also investigated in many disperse literatures \citep{raiteri03,papadakis03,hu06,wu06,wu11,dai09,zhang10a,dai11,fan14}.

BWB and RWB trends are two common well-observed behaviors in blazars.
However,  some complicated color phenomena have also been found in some cases.
In general, there are five kinds of color behaviors: BWB in the whole data sets; RWB in the whole data sets; cycles or loop-like pattern; RWB at the low state while BWB at the high state (RWB to BWB); stable-when-brighter (SWB) or no correlation with brightness in the whole  data sets (see \cite{zhang14} and references therein).
Up to now, FSRQs and BL Lacs seem not to have universal features in color variations.
The purpose of this paper is to investigate their color behaviors using a much larger sample of blazars with almost simultaneous measurements, and expect to provide results with higher confidence levels.  The paper is organized as following: in Section 2, the infrared and optical data as well as their variations are described; in Section 3, the color behavior is studied; then the discussion and conclusions are given in Section 4.

\begin{table}
\begin{center}
\caption[]{The Property of Variations of Optical and Infrared Data.}
\label{table:statistics}
\tiny

 \begin{tabular}{llp{0.5cm}cccccccccc}
  \hline\noalign{\smallskip}
  $Object$ & $Class$ & $\Delta T$ (day) & $N$  &  $J_{Fvar}$ &  $R_{Fvar}$ & $r_{J-R}$ & $Prob.$ & $r_{C-R}$ & $Prob.$ & $r_{C-J}$ & $Prob.$ & $Behavior$ \\
  \hline\noalign{\smallskip}
PMN 0017-0512   &FSRQ   &211   &73  &0.82$\pm$0.07 &0.67$\pm$0.06 &0.97 &$<10^{-9}$         &0.67 &$<10^{-11}$ &0.82   &$<10^{-9}$               &RWB  \\
PKS 0035-252    &FSRQ   &357   &9   &0.43$\pm$0.11 &0.38$\pm$0.10 &0.11 &0.78               &-0.49&0.18                &0.81 &$8.0\times10^{-3}$ &RWB\\
PKS 0208-512    &FSRQ   &2016  &381 &0.96$\pm$0.03 &0.86$\pm$0.03 &0.96 &$<10^{-9}$         &0.30 &$2.7\times10^{-9}$  &0.55 &$<10^{-9}$        &RWB\\
4C+01.02        &FSRQ   &76    &8   &0.32$\pm$0.08 &0.17$\pm$0.04 &0.73 &0.04               &0.34 &0.40                &0.89 &$2.7\times10^{-3}$ &RWB\\
AO 0235+164     &FSRQ   &1816  &221 &1.57$\pm$0.07 &1.59$\pm$0.08 &0.98 &$<10^{-9}$         &0.21 &$2.1\times10^{-3}$  &0.38 &$4.2\times10^{-9}$ &RWB\\
PKS 0235-618    &FSRQ   &158   &14  &0.25$\pm$0.05 &0.32$\pm$0.06 &0.48 &0.09               &-0.72&$3.9\times10^{-3}$  &0.27 &0.35               &BWB \\
1RXS 0238-3116  &BL Lac &654   &20  &0.20$\pm$0.03 &0.29$\pm$0.05 &-0.10&0.68               &-0.87&$7.7\times10^{-7}$  &0.58 &$7.2\times10^{-3}$ &BWB(RWB)\\
PKS 0250-225    &FSRQ   &90    &50  &0.40$\pm$0.04 &0.40$\pm$0.04 &0.85 &$<10^{-9}$         &-0.26&0.07                &0.29 &0.04               &None\\
PKS 0301-243    &BL Lac &537   &48  &0.15$\pm$0.02 &0.18$\pm$0.02 &0.84 &$<10^{-9}$         &-0.56&$4.1\times10^{-5}$  &-0.02&0.89               &BWB\\
BG6 J0316+0904  &BL Lac &783   &12  &0.16$\pm$0.03 &0.25$\pm$0.05 &0.92 &$2.3\times10^{-5}$ &-0.82&$9.7\times10^{-4}$  &-0.54&0.07               &BWB\\
PKS 0402-362    &FSRQ   &863   &75  &0.26$\pm$0.02 &0.17$\pm$0.01 &0.86 &$<10^{-9}$         &0.34 &$2.5\times10^{-3}$  &0.78 &$<10^{-9}$         &RWB\\
PMN 0413-5332   &FSRQ   &763   &11  &0.48$\pm$0.12 &0.10$\pm$0.06 &-0.26&0.44               &-0.49&0.12                &0.97 &$9.0\times10^{-7}$ &RWB\\
PKS 0422+004    &BL Lac &498   &11  &0.42$\pm$0.09 &0.44$\pm$0.09 &0.99 &$1.5\times10^{-8}$ &-0.28&0.40                &-0.13&0.71               &None\\
PKS 0426-380    &FSRQ   &419   &163 &0.35$\pm$0.02 &0.34$\pm$0.02 &0.92 &$<10^{-9}$         &-0.18&0.02                &0.22 &$4.9\times10^{-3}$ &RWB\\
NRAO 190        &FSRQ   &236   &15  &0.86$\pm$0.16 &0.77$\pm$0.14 &0.94 &$1.7\times10^{-7}$ &0.32 &0.24                &0.62 &0.01               &RWB\\
PKS 0454-234    &FSRQ   &847   &286 &0.61$\pm$0.03 &0.68$\pm$0.03 &0.99 &$<10^{-9}$         &-0.62&$<10^{-9}$          &-0.48&$<10^{-9}$         &BWB\\
PKS 0454-46     &FSRQ   &1193  &47  &0.14$\pm$0.02 &0.15$\pm$0.02 &0.21 &0.16               &-0.61&$5.6\times10^{-6}$  &0.65 &$8.5\times10^{-7}$ &RWB(BWB)\\
S3 0458-02      &FSRQ   &496   &39  &0.34$\pm$0.04 &0.31$\pm$0.04 &0.93 &$<10^{-9}$         &0.12 &0.45                &0.48 &$1.9\times10^{-3}$ &RWB\\
PKS 0502+049    &FSRQ   &309   &36  &0.52$\pm$0.06 &0.34$\pm$0.04 &0.76 &$9.4\times10^{-8}$ &0.01 &0.97                &0.66 &$1.2\times10^{-5}$ &RWB\\
PMN 0507-6104   &FSRQ   &816   &6   &0.61$\pm$0.18 &0.66$\pm$0.19 &0.75 &0.09               &-0.25&0.64                &0.46 &0.36               &None\\
PKS 0528+134    &FSRQ   &2091  &160 &0.54$\pm$0.03 &0.31$\pm$0.02 &0.78 &$<10^{-9}$         &0.23 &$4.1\times10^{-3}$  &0.79 &$<10^{-9}$         &RWB\\
PMN J0531-4827  &$-$    &1215  &175 &0.94$\pm$0.05 &1.01$\pm$0.05 &0.98 &$<10^{-9}$         &-0.34&$3.6\times10^{-6}$  &-0.13&0.09               &BWB\\
PKS 0537-441    &FSRQ   &1115  &141 &0.77$\pm$0.05 &0.66$\pm$0.04 &0.97 &$<10^{-9}$         &0.16 &0.05                &0.40 &$1.1\times10^{-6}$ &RWB\\
PMN J0623+3350  &$-$    &61    &20  &0.40$\pm$0.07 &0.54$\pm$0.09 &0.75 &$1.2\times10^{-4}$ &-0.67&$1.2\times10^{-3}$  &-0.02&0.93               &BWB\\
J0630.9-2406    &BL Lac &827   &18  &0.21$\pm$0.04 &0.12$\pm$0.02 &0.37 &0.14               &-0.26&0.30                &0.81 &$5.5\times10^{-5}$ & RWB\\
PKS 0637-75     &FSRQ   &1193  &72  &0.09$\pm$0.01 &0.07$\pm$0.01 &0.30 &0.01               &-0.46&$5.7\times10^{-5}$  &0.72 &$<10^{-9}$         &RWB(BWB)\\
PKS 0727-11     &FSRQ   &989   &18  &0.64$\pm$0.12 &0.88$\pm$0.15 &0.96 &$<10^{-9}$         &-0.86&$4.1\times10^{-6}$  &-0.69&$1.7\times10^{-3}$ &BWB\\
PMN J0816-1311  &BL Lac &843   &21  &0.22$\pm$0.03 &0.23$\pm$0.04 &0.97 &$<10^{-9}$         &-0.36&0.10                &-0.11&0.62               &None\\
PKS 0818-128    &BL Lac &826   &20  &0.23$\pm$0.04 &0.36$\pm$0.06 &0.53 &0.02               &-0.76&$1.1\times10^{-4}$  &0.16 &0.51               &BWB\\
BZQ J0850-1213  &FSRQ   &1206  &97  &0.55$\pm$0.04 &0.53$\pm$0.04 &0.91 &$<10^{-9}$         &-0.21&0.04                &0.21 &0.04               &None\\
OJ 287          &BL Lac &2203  &364 &0.42$\pm$0.02 &0.41$\pm$0.02 &0.92 &$<10^{-9}$         &-0.04&0.49                &0.35 &$<10^{-9}$         &RWB\\
PKS 1004-217    &FSRQ   &1087  &81  &0.20$\pm$0.02 &0.14$\pm$0.01 &-0.12&0.29               &-0.72&$<10^{-9}$          &0.77 &$<10^{-9}$         &RWB(BWB)\\
PKS 1056-113    &BL Lac &728   &43  &0.37$\pm$0.04 &0.43$\pm$0.05 &0.97 &$<10^{-9}$         &-0.35&0.02                &-0.10&0.52               &None\\
PKS 1124-186    &FSRQ   &351   &34  &0.34$\pm$0.04 &0.34$\pm$0.04 &1.00 &$<10^{-9}$         &0.54 &$9.5\times10^{-4}$  &0.60 &$2.0\times10^{-4}$ &RWB\\
PKS 1127-14     &FSRQ   &1103  &37  &0.21$\pm$0.03 &0.17$\pm$0.02 &-0.12&0.49               &-0.62&$4.6\times10^{-5}$  &0.85 &$<10^{-9}$         &RWB(BWB)\\
PKS 1144-379    &FSRQ   &1115  &128 &0.56$\pm$0.04 &0.65$\pm$0.04 &0.93 &$<10^{-9}$         &-0.66&$<10^{-9}$          &-0.34&$6.9\times10^{-5}$ &BWB\\
QSO B1212+078   &BL Lac &942   &9   &0.07$\pm$0.02 &0.12$\pm$0.03 &0.47 &0.21               &-0.78&0.01                &0.18 &0.63               &BWB\\
3C 273          &FSRQ   &2156  &287 &0.06$\pm$0.00 &0.04$\pm$0.00 &0.14 &0.02               &-0.54&$<10^{-9}$          &0.76 &$<10^{-9}$         &RWB(BWB)\\
PKS 1244-255    &FSRQ   &1032  &107 &0.55$\pm$0.04 &0.48$\pm$0.03 &0.98 &$<10^{-9}$         &0.69 &$<10^{-9}$          &0.81 &$<10^{-9}$         &RWB\\
3C 279          &FSRQ   &2207  &447 &0.68$\pm$0.02 &0.72$\pm$0.02 &0.99 &$<10^{-9}$         &-0.46&$<10^{-9}$          &-0.33&$<10^{-9}$         &BWB\\
PKS 1329-049    &FSRQ   &376   &11  &0.24$\pm$0.08 &0.48$\pm$0.11 &0.08 &0.82               &-0.73&0.01                &0.62 &0.04               &BWB\\
PKS 1335-127    &FSRQ   &1077  &54  &0.91$\pm$0.09 &0.87$\pm$0.08 &0.95 &$<10^{-9}$         &0.10 &0.47                &0.42 &$1.6\times10^{-3}$ &RWB\\
PKS B1406-076   &FSRQ   &2201  &180 &0.61$\pm$0.03 &0.44$\pm$0.02 &0.80 &$<10^{-9}$         &0.14 &0.06                &0.71 &$<10^{-9}$         &RWB\\
PKS B1424-418   &FSRQ   &765   &278 &0.71$\pm$0.03 &0.74$\pm$0.03 &0.99 &$<10^{-9}$         &-0.38&$<10^{-9}$          &-0.26&$1.3\times10^{-5}$ &BWB\\
PKS 1508-05     &FSRQ   &745   &25  &0.20$\pm$0.03 &0.10$\pm$0.02 &0.25 &0.22               &-0.27&0.20                &0.86 &$2.7\times10^{-8}$ &RWB\\

\noalign{\smallskip}\hline
\end{tabular}
\end{center}
\end{table}

\begin{table}[t]
\begin{center}
\addtocounter{table}{-1}
\caption[]{Continued.}
\label{table:statistics1}
\tiny
  \begin{tabular}{llp{0.5cm}ccccccccccc}
  \hline\noalign{\smallskip}
  $Object$ & $Class$ & $\Delta T$ (day) & $N$  &  $J_{Fvar}$ &  $R_{Fvar}$ & $r_{J-R}$ & $Prob.$ & $r_{C-R}$ & $Prob.$ & $r_{C-J}$ & $Prob.$ & $Behavior$ \\
  \hline\noalign{\smallskip}
PKS 1510-089   &FSRQ    &2207  &405 &0.96$\pm$0.03 &0.85$\pm$0.03 &0.96 &$<10^{-9}$         &0.53 &$<10^{-9}$         &0.74 &$<10^{-9}$         &RWB\\
PKS 1514-241   &BL Lac  &1065  &68  &0.15$\pm$0.01 &0.15$\pm$0.01 &0.81 &$<10^{-9}$         &-0.27&0.03               &0.35 &$3.1\times10^{-3}$ &RWB\\
PKS 1550-242   &$-$     &304   &17  &0.55$\pm$0.10 &0.57$\pm$0.10 &0.97 &$<10^{-9}$         &-0.23&0.38               &0.00 &1.00               &None\\
PMN J1610-6649 &BL Lac  &741   &16  &0.19$\pm$0.03 &0.16$\pm$0.03 &0.97 &$<10^{-9}$         &0.53 &0.03               &0.72 &$1.8\times10^{-3}$ &RWB\\
PKS 1622-253   &FSRQ    &988   &6   &0.46$\pm$0.13 &0.48$\pm$0.15 &0.74 &0.09               &-0.03&0.95               &0.65 &0.16               &RWB\\
PKS 1622-297   &FSRQ    &2165  &221 &0.54$\pm$0.03 &0.37$\pm$0.02 &0.85 &$<10^{-9}$         &0.27 &$4.9\times10^{-5}$ &0.74 &$<10^{-9}$         &RWB\\
1717-5156      &$-$     &455   &133 &0.88$\pm$0.05 &0.39$\pm$0.02 &0.86 &$<10^{-9}$         &0.59 &$<10^{-9}$         &0.92 &$<10^{-9}$         &RWB\\
PKS 1730-130   &FSRQ    &1985  &243 &0.41$\pm$0.02 &0.32$\pm$0.01 &0.79 &$<10^{-9}$         &-0.05&0.39               &0.57 &$<10^{-9}$         &RWB\\
OT 081         &BL Lac  &379   &14  &0.81$\pm$0.15 &0.85$\pm$0.16 &0.99 &$<10^{-9}$         &-0.33&0.25               &-0.19&0.51               &None\\
PMN J1921-1607 &BL Lac  &543   &13  &0.28$\pm$0.06 &0.23$\pm$0.05 &0.96 &$2.3\times10^{-7}$ &0.63 &0.02               &0.82 &$5.6\times10^{-4}$ &RWB\\
PKS 1954-388   &FSRQ    &1118  &54  &0.42$\pm$0.04 &0.29$\pm$0.03 &0.96 &$<10^{-9}$         &0.72 &$<10^{-9}$         &0.88 &$<10^{-9}$         &RWB\\
PKS 1958-179   &FSRQ    &919   &18  &1.09$\pm$0.18 &1.17$\pm$0.20 &0.98 &$<10^{-9}$         &-0.59&0.01               &-0.42&0.08               &BWB\\
PKS 2052-474   &FSRQ    &1553  &176 &0.80$\pm$0.04 &0.77$\pm$0.04 &0.96 &$<10^{-9}$         &0.10 &0.18               &0.39 &$7.8\times10^{-8}$ &RWB\\
2FGL J2055.4   &BL Lac  &752   &11  &0.15$\pm$0.04 &0.21$\pm$0.05 &-0.29&0.38               &-0.83&$1.4\times10^{-3}$ &0.77 &$5.4\times10^{-3}$ &BWB(RWB)\\
-0023 & & & & & & & &&\\
PKS 2142-75    &FSRQ    &1316  &157 &0.52$\pm$0.03 &0.17$\pm$0.01 &0.88 &$<10^{-9}$         &0.77 &$<10^{-9}$         &0.98 &$<10^{-9}$         &RWB\\
PKS 2149-306   &FSRQ    &158   &22  &0.17$\pm$0.03 &0.09$\pm$0.01 &0.92 &$2.0\times10^{-9}$ &0.72 &$1.7\times10^{-4}$ &0.94 &$<10^{-9}$         &RWB\\
PKS 2155-304   &BL Lac  &1940  &363 &0.38$\pm$0.01 &0.37$\pm$0.01 &0.99 &$<10^{-9}$         &0.06 &0.29               &0.22 &$1.7\times10^{-5}$ &RWB\\
3C 446         &BL Lac  &1111  &56  &0.34$\pm$0.04 &0.50$\pm$0.05 &0.35 &$7.3\times10^{-3}$ &-0.60&$9.8\times10^{-7}$ &0.53 &$2.2\times10^{-5}$ &BWB(RWB)\\
PKS 2227-08    &FSRQ    &915   &43  &0.28$\pm$0.03 &0.51$\pm$0.06 &0.32 &0.04               &-0.87&$<10^{-9}$         &0.18 &0.24               &BWB\\
PKS 2232-488   &FSRQ    &661   &10  &0.41$\pm$0.09 &0.26$\pm$0.06 &0.66 &0.04               &0.02 &0.96               &0.76 &0.01               &RWB\\
PKS 2233-148   &BL Lac  &507   &122 &0.76$\pm$0.05 &0.83$\pm$0.05 &1.00 &$<10^{-9}$         &-0.53&$<10^{-9}$         &-0.45&$2.6\times10^{-7}$ &BWB\\
PKS 2240-260   &BL Lac  &1147  &20  &0.24$\pm$0.04 &0.22$\pm$0.04 &0.92 &$1.1\times10^{-8}$ &0.22 &0.35               &0.59 &$6.6\times10^{-3}$ &RWB\\
3C 454.3       &FSRQ    &1985  &505 &0.90$\pm$0.03 &0.81$\pm$0.03 &0.99 &$<10^{-9}$         &0.76 &$<10^{-9}$         &0.85 &$<10^{-9}$         &RWB\\
PKS 2255-282   &FSRQ    &947   &42  &0.57$\pm$0.06 &0.49$\pm$0.05 &0.94 &$<10^{-9}$         &0.26 &0.09               &0.58 &$6.6\times10^{-5}$ &RWB\\
  1ES 2322-409 &BL Lac  &809   &19  &0.20$\pm$0.03 &0.23$\pm$0.04 &0.96 &$<10^{-9}$         &-0.58&$9.1\times10^{-3}$ &-0.33&0.17               &BWB\\
PKS 2326-502   &FSRQ    &128   &76  &0.58$\pm$0.05 &0.72$\pm$0.06 &0.97 &$<10^{-9}$         &-0.56&$1.8\times10^{-7}$ &-0.35&$2.2\times10^{-3}$ &BWB\\
PMN 2345-1555  &FSRQ    &790   &167 &0.57$\pm$0.03 &0.63$\pm$0.03 &0.98 &$<10^{-9}$         &-0.40&$1.1\times10^{-7}$ &-0.19&0.01               &BWB\\
PKS 2345-16    &FSRQ    &372   &22  &0.81$\pm$0.12 &0.79$\pm$0.12 &-0.97&$<10^{-9}$         &-0.99&$<10^{-9}$         &0.99 &$<10^{-9}$         &RWB(BWB)\\
H 2356-309     &BL Lac  &478   &10  &0.09$\pm$0.02 &0.18$\pm$0.04 &0.69 &0.03               &-0.83&$3.1\times10^{-3}$ &-0.16&0.65               &BWB\\
4C+01.28       &BL Lac  &27    &8   &0.35$\pm$0.09 &0.29$\pm$0.07 &-0.95&$3.5\times10^{-4}$ &-0.99&$4.6\times10^{-6}$ &0.99 &$7.2\times10^{-6}$ &BWB(RWB)\\
\noalign{\smallskip}\hline
\end{tabular}
\end{center}
\end{table}

\section{Infrared and optical data \& variability}

In order to investigate the statistics of the color behavior of blazars, we collect 71 blazars as well as 4 possible blazars from SMARTS sources.
SMARTS program employs four meter-class telescopes (1.5-m, 1.3-m, 1.0-m and 0.9-m) of the Small and Moderate Aperture Research Telescope System (SMARTS) at Cerro Tololo Interamerican Observatory (CTIO) in Chile to observe the bright southern gamma-ray blazars on a regular cadence, at both optical and near-infrared wavelengths.
Data reduction and analysis of SMARTS data have been described in \cite{bonning12}. Table~\ref{table:statistics} lists our sample which comprises 49 FSRQs, 22 BL Lacs and 4 unidentified type sources.  The first column gives the object name, the second column gives  the spectral classification, the third one gives the monitoring duration in days, the fourth one gives the number of data points for color index calculation. Most of the sources are observed in $B$, $V$, $R$, $J$ and $K$ bands. On the whole, there are the most data points in $R$ and $J$ bands. So, these two bands have been selected to study color variations. However, the observations for each source are not uniform. Some objects have been observed for more than 2000 days, while some others have been observed for less than 100 days. Some sources have been well observed with hundreds of data points and some less than 10 data points.

\begin{figure}
\centering
\includegraphics[width=8.5cm]{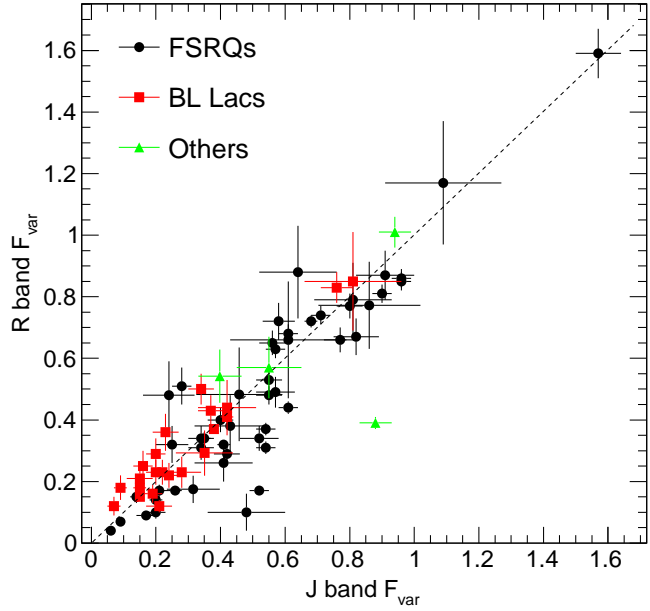}
 \caption{$F_{var}$s of $R$ band versus those of $J$ band. The dots, squares and triangles represent FSRQs, BL Lacs and unidentified type objects, respectively. The dashed line means that the $F_{var}$ of $J$ band equal to that of $R$ band.}
 \label{fvar-r-j}
\end{figure}

In order to compare the amplitude of the variations in different blazars, and in different bands for the same blazar, we compute the fractional variability amplitude $F_{var}$ \citep{vaughan03}, which represents the average amplitude of the observed variations as a
percentage of the light curve mean. The fractional variability amplitude is defined as
$F_{var}=\sqrt{(S^{2}-<\varepsilon^{2}_{err}>)/{<x>}^{2}},$
where $S^{2}$ is the sample variance of the light curve,
$<x>$ is the average flux and $<\varepsilon^{2}_{err}>$ is the mean of the squared
measurement uncertainties.  The $F_{var}$ values of $J$ and $R$ bands are listed
in the fifth and sixth columns of Table~\ref{table:statistics}. In Fig.~\ref{fvar-r-j} we plot the $F_{var}$ of $R$ band versus that of $J$ band for each object. It can be seen that the objects show great differences in variations. Most of the objects show significant variabilities. Several sources display extreme variability, such as AO 0235-164, with $F_{var}$$ = $1.57 in $J$ band and $F_{var}$$ = $1.59 in $R$ band. However, there are also a few sources showing weak variabilities, such as 3C 273, with $F_{var}$ = 0.06 in $J$ band and $F_{var}$ = 0.04 in $R$ band.

\begin{figure}[h]
  \begin{minipage}[t]{0.495\linewidth}
  \centering
   \includegraphics[width=59mm,height=58mm]{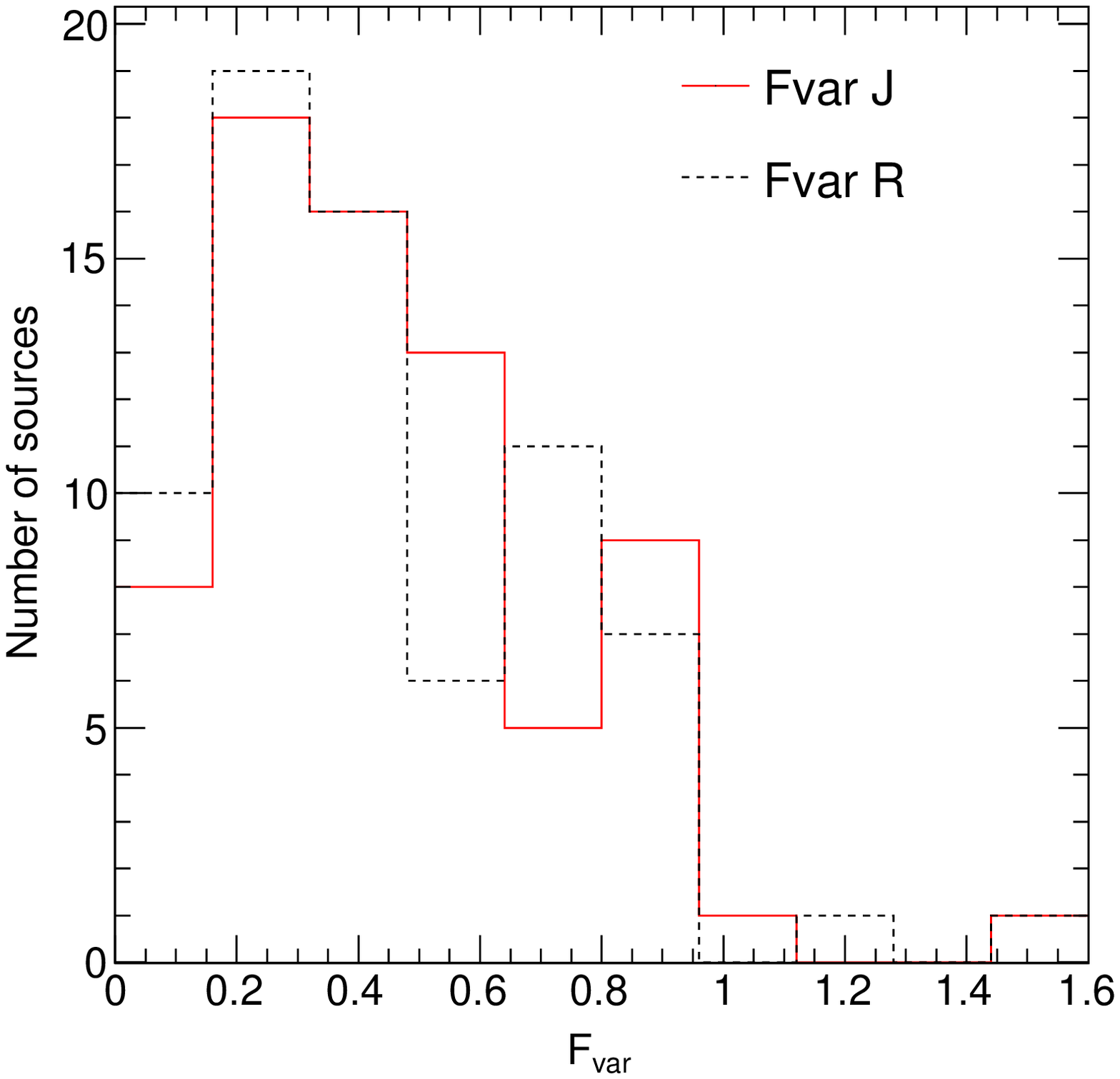}
   \caption{{\small Distributions of $F_{vars}$ of the optical $R$ band (the dashed line) and the infrared $J$ band (the solid line).} }
    \label{dis-fvar}
  \end{minipage}%
  \begin{minipage}[t]{0.495\textwidth}
  \centering
   \includegraphics[width=59mm,height=58mm]{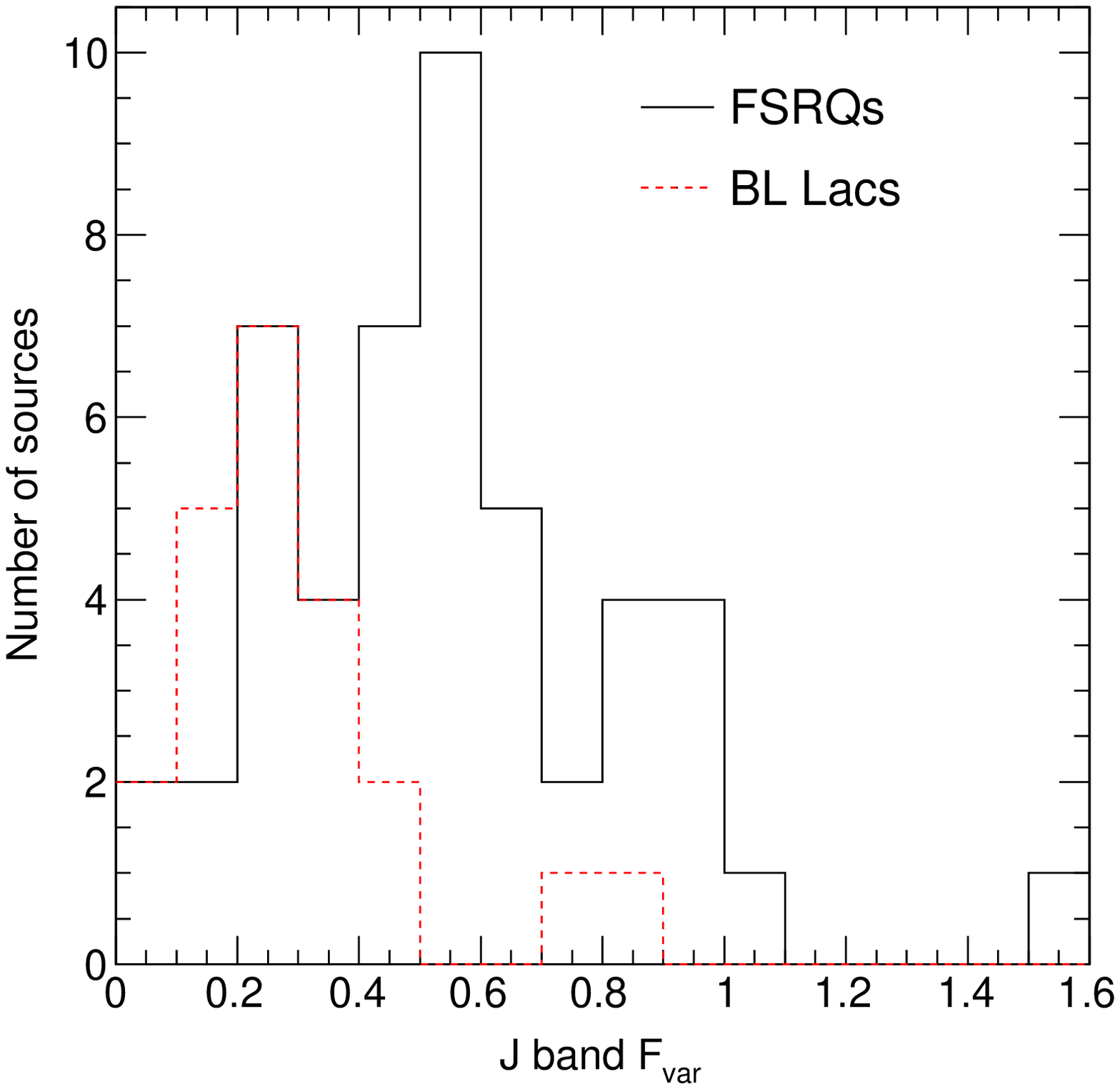}
  \caption{{\small Distributions of $J$ band $F_{var}s$ of FSRQs (the solid line) and BL Lacs (the dashed line).}}
   \label{dis-fvar-f-b}
  \end{minipage}%
\end{figure}

Fig.~\ref{dis-fvar} plots the distribution of $F_{var}s$ of $R$ band and $J$ band. On the whole,  $F_{var}s$ of $R$ band have no significant difference with those of $J$ band for all objects, which means that the variability amplitudes of $R$ and $J$ band in blazars are comparable. \cite{ghosh00}  suggested that it may not be correct to generalize that the amplitude of the variation in blazars is systematically larger at higher frequency. However, as can be seen in Fig.~\ref{fvar-r-j}, $F_{var}$s of $J$ band are generally larger than those of $R$ band for most of FSRQs (with $prob.$ = 0.025), while $F_{var}$s of $J$ band are, in general, smaller than those of $R$ band for most of BL Lacs (with $prob.$ = 0.033).  \cite{bonning12} also found that 8 out of 9 FSRQs are more variable in $J$ band than $B$ band.
 As can also be seen in Fig.~\ref{fvar-r-j}, the $F_{var}$s of FSRQs seem to be larger than those of BL Lacs. Fig.~\ref{dis-fvar-f-b} displays  the $F_{var}$ distribution of FSRQs and BL Lacs  in $J$ band.
 One can clearly see that
  $J$ band $F_{var}s$ of FSRQs are larger than those of BL Lacs. The difference is significant at a level of $5.7\times10^{-4}$.

We investigate the correlation between $R$ and $J$ bands. The $R$ band versus $J$ band is plotted in the left panels in Fig.~\ref{color}. Linear fitting is carried out by the least square method (the solid line in the figure).  In order to quantitatively search for the relation between these two bands, we calculate the correlation coefficient and estimate the significance level, and list them in Columns 7 and 8 of Table~\ref{table:statistics}. From Fig.~\ref{color} and the correlation coefficient, one can see that blazars show well-correlated variations between $R$ and $J$ bands, except for several sources showing weak or no, even negative correlations.

\section{Color variations}

The color variability can reflect the spectral variability. Since these sources are highly variable, simultaneous or quasi-simultaneous observations are essential for color index calculation. The time series in the optical $R$ band and the infrared $J$ band are quasi-simultaneous and the best sampled  in the SMARTS data. This provides us a chance to study in detail the spectral variations of the large sample of blazars on long time scales. $J-R$ color index is calculated pair by pair between $R$ and $J$ bands within the time interval of 15 minutes for each object. More than 90 percent of intervals are less than 5 minutes. Very few extremely strange data points have been discarded. The middle and right panels of Fig.~\ref{color} show the $J-R$ color variations with the brightness of $R$ band and $J$ band, respectively.  To obtain the relation between the color index and the source brightness, we calculate the correlation coefficient $r$, and estimate the significance level. If $r$ $>=$ 0.2 and the significance level $p<=$ 0.01, it means somewhat a good correlation between color index and magnitude. It suggests that the source becomes redder when the source becomes brighter.  If $r$ $<=$ $-$0.2 and the significance level $p<=$  0.01, it means a well negative correlation, and suggests that the source tends to be bluer when the source is brighter. The other cases mean there is no correlation between color index and source brightness. Column 9 of Table~\ref{table:statistics} lists the correlation coefficient between the color index and the magnitude of $R$ band (denoted by $r_{C-R}$). Its significance level is estimated and listed in Column 10.
The correlation coefficient between the color index and the magnitude of $J$ band ($r_{C-J}$) and its significance level are also calculated and listed in Columns 11 and 12.
Overall, 43 sources (35 FSRQs, 7 BL Lacs and 1 unidentified object) show RWB trends and 24 sources (11 FSRQs, 11 BL Lacs and 2 unidentified objects) show BWB trends. There are 10 sources (6 FSRQs and 4 BL Lacs) showing BWB trends in the $Color-R$ plane and  opposite trends  in the $Color-J$ plane. The color behaviors are listed in Column 13 of Table~\ref{table:statistics}.

\begin{figure}[b]
\centering
 \includegraphics[width=10.5cm]{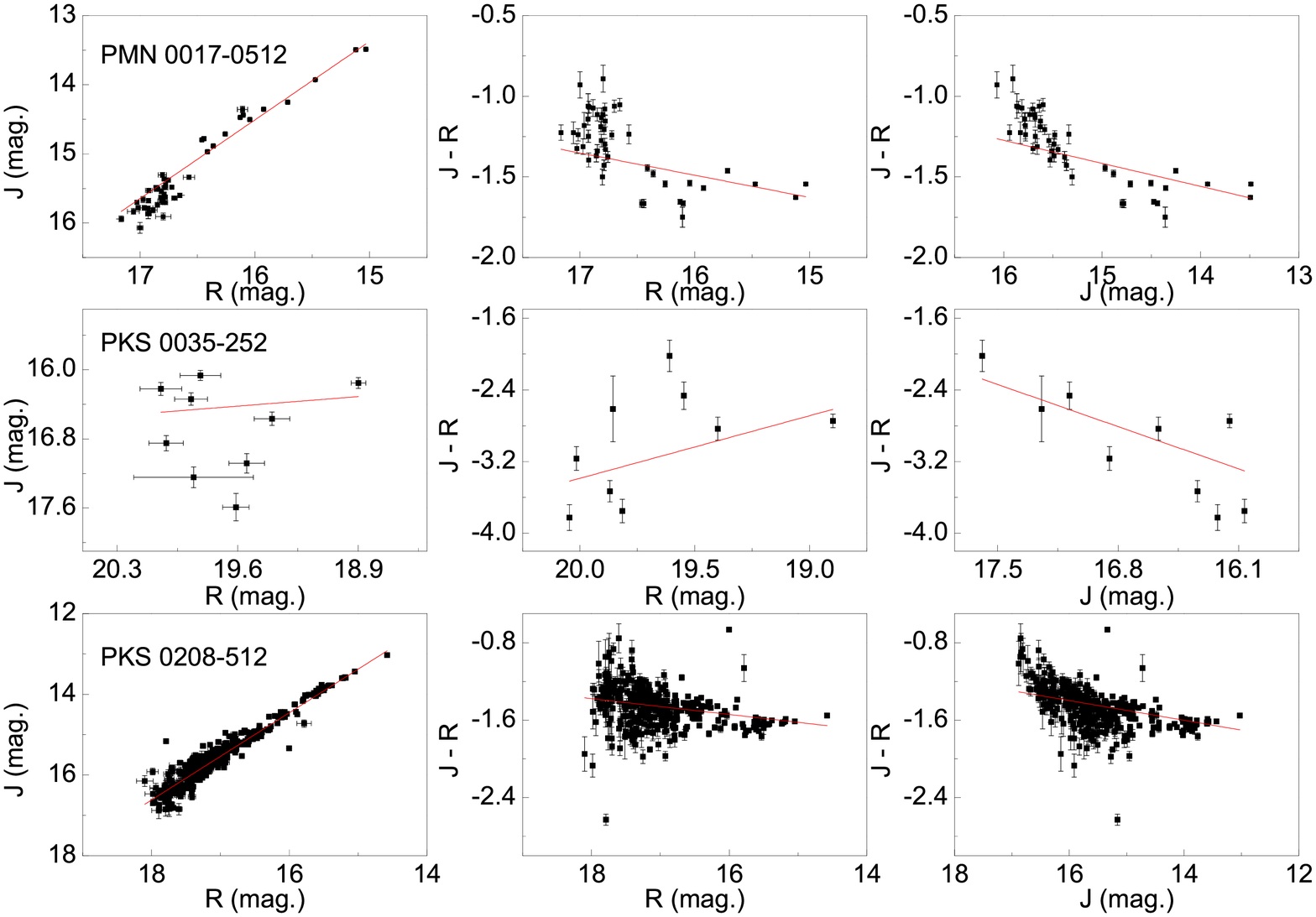}
 \caption{Left: Relations between the $J$ and $R$ magnitudes. Middle: Color index versus the magnitude of $R$ band. Right: Color index versus the magnitude of $J$ band. The solid lines represent the best linear fitting.}
 \label{color}
\end{figure}

\addtocounter{figure}{-1}
\begin{figure}
\centering
\includegraphics[width=6.6cm]{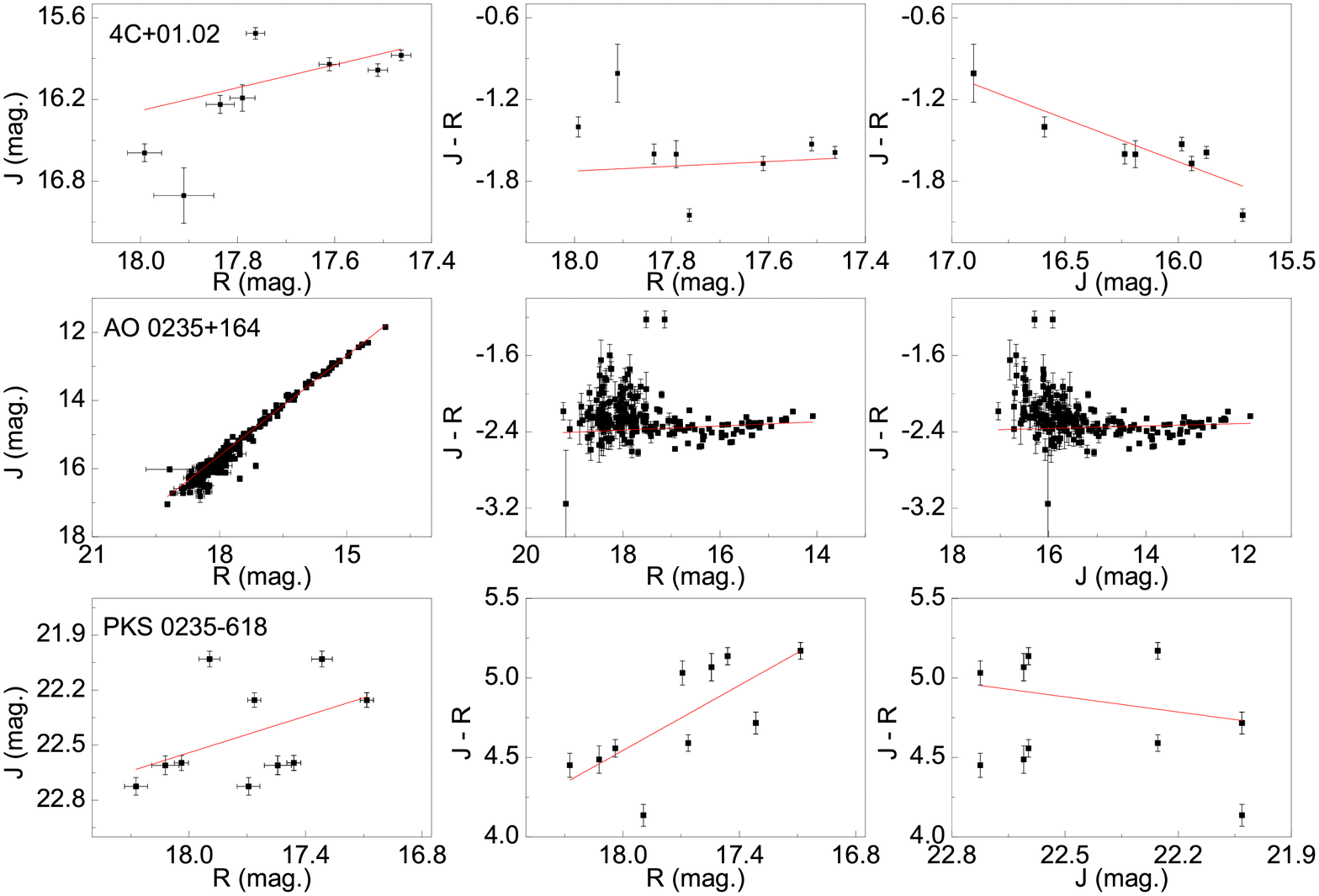}
 \includegraphics[width=6.3cm]{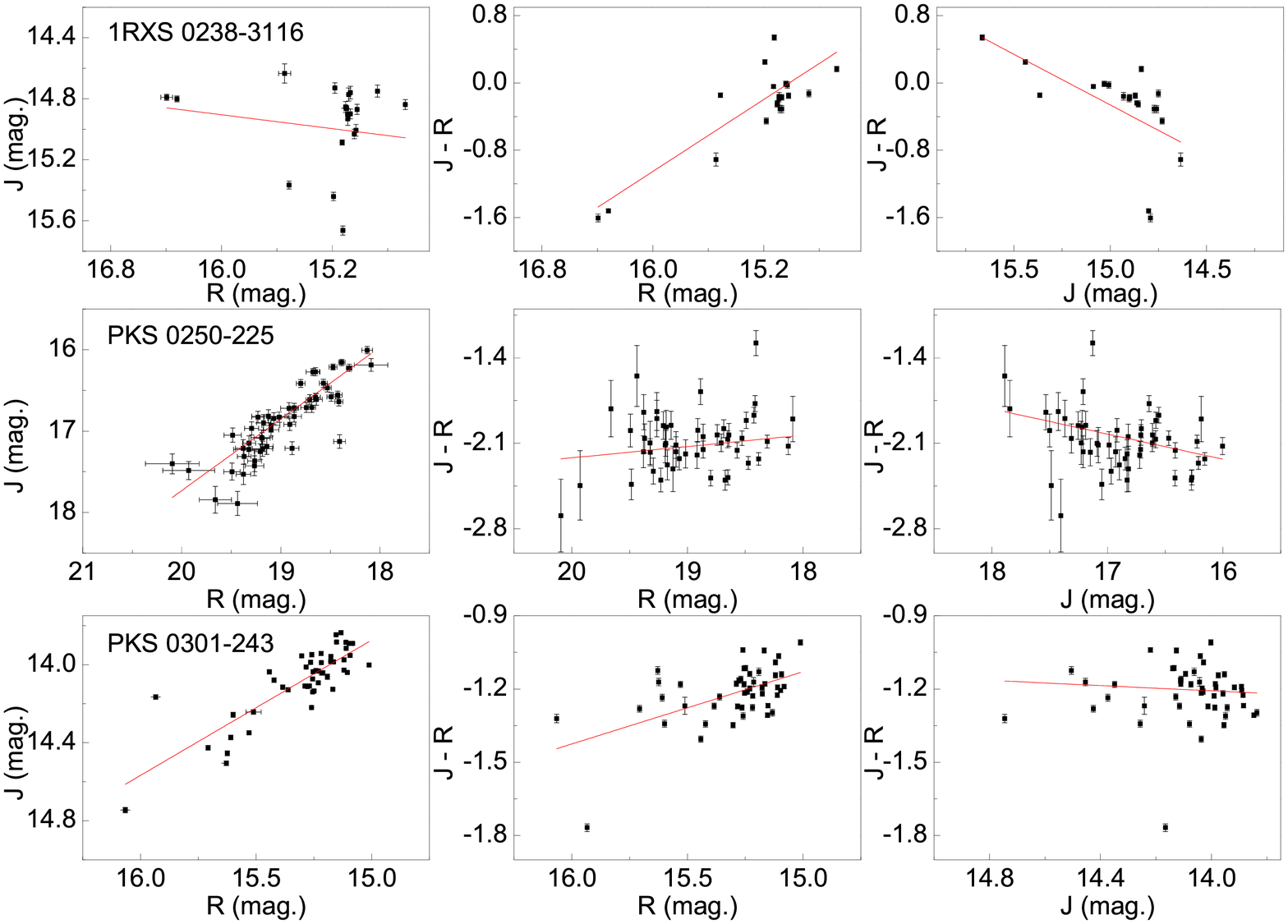}
 \includegraphics[width=6.5cm]{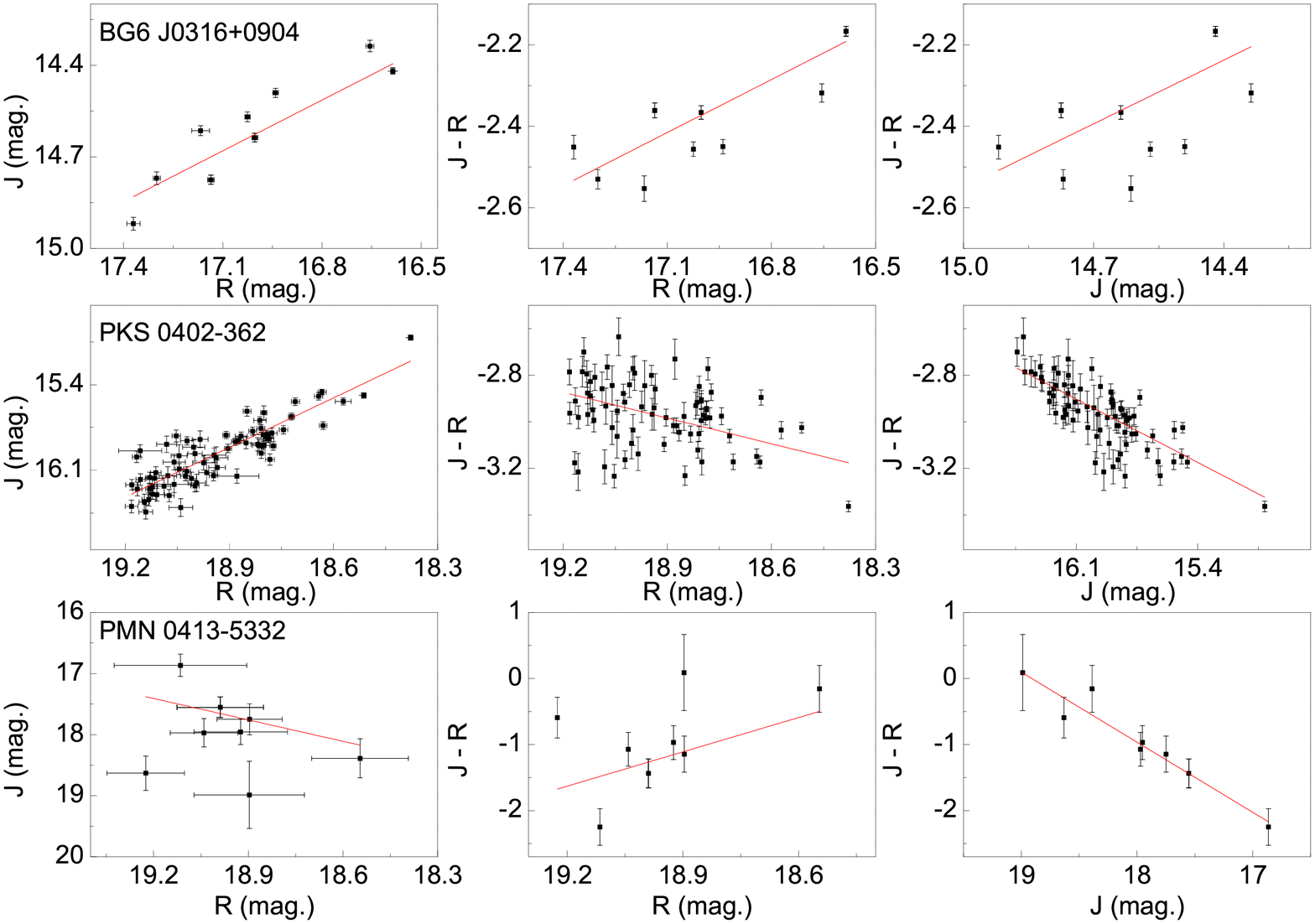}
  \includegraphics[width=6.5cm]{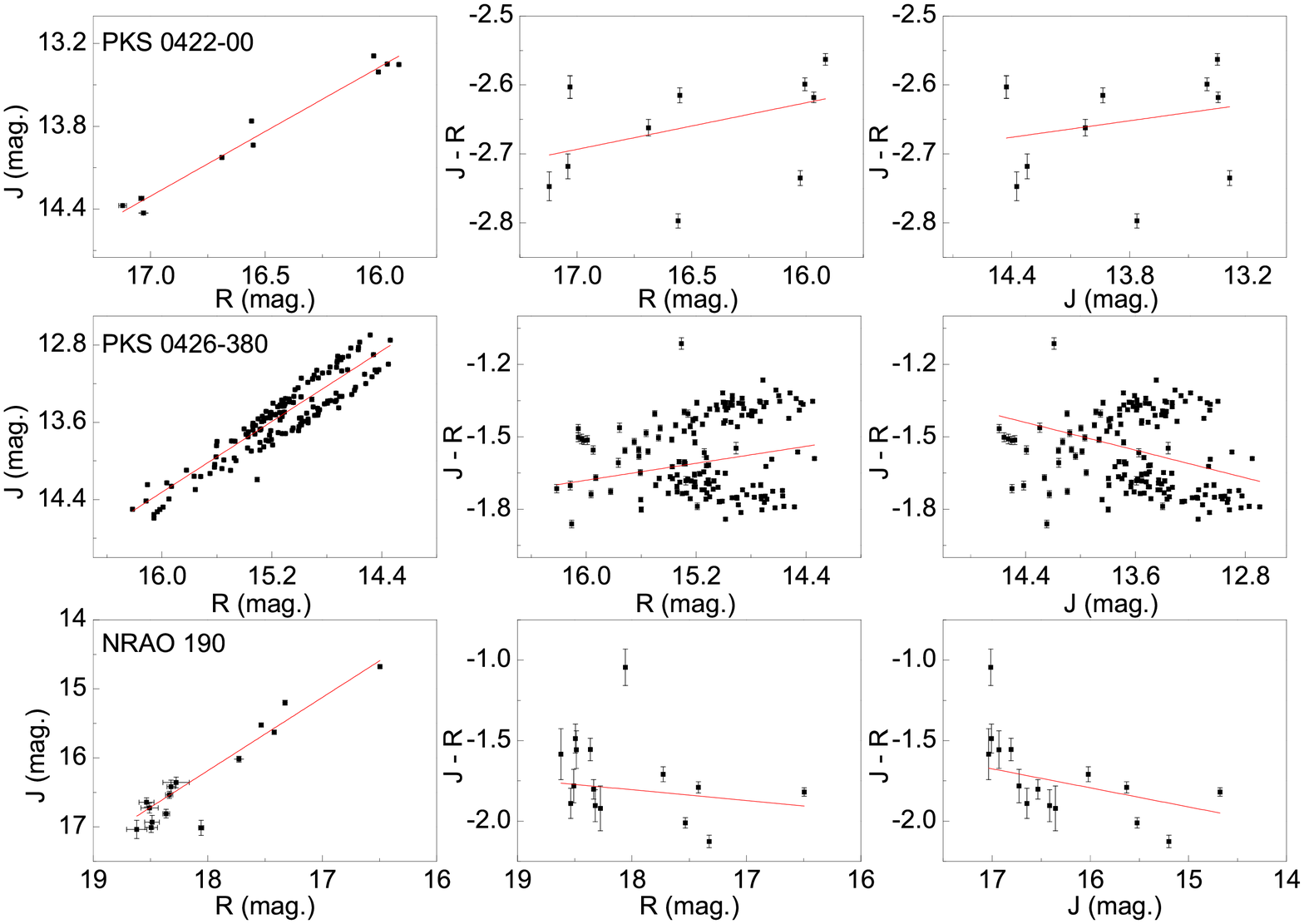}
  \includegraphics[width=6.5cm]{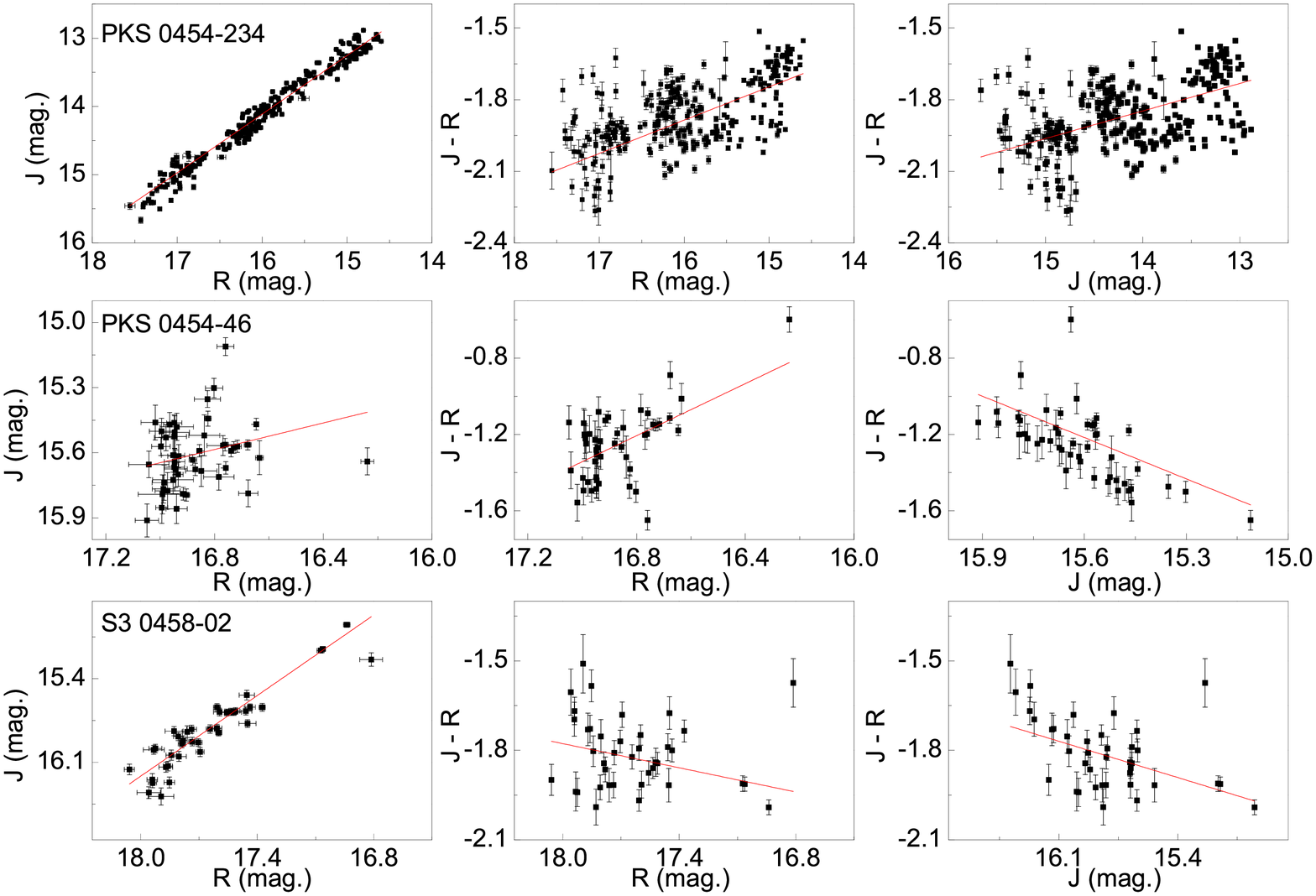}
  \includegraphics[width=6.5cm]{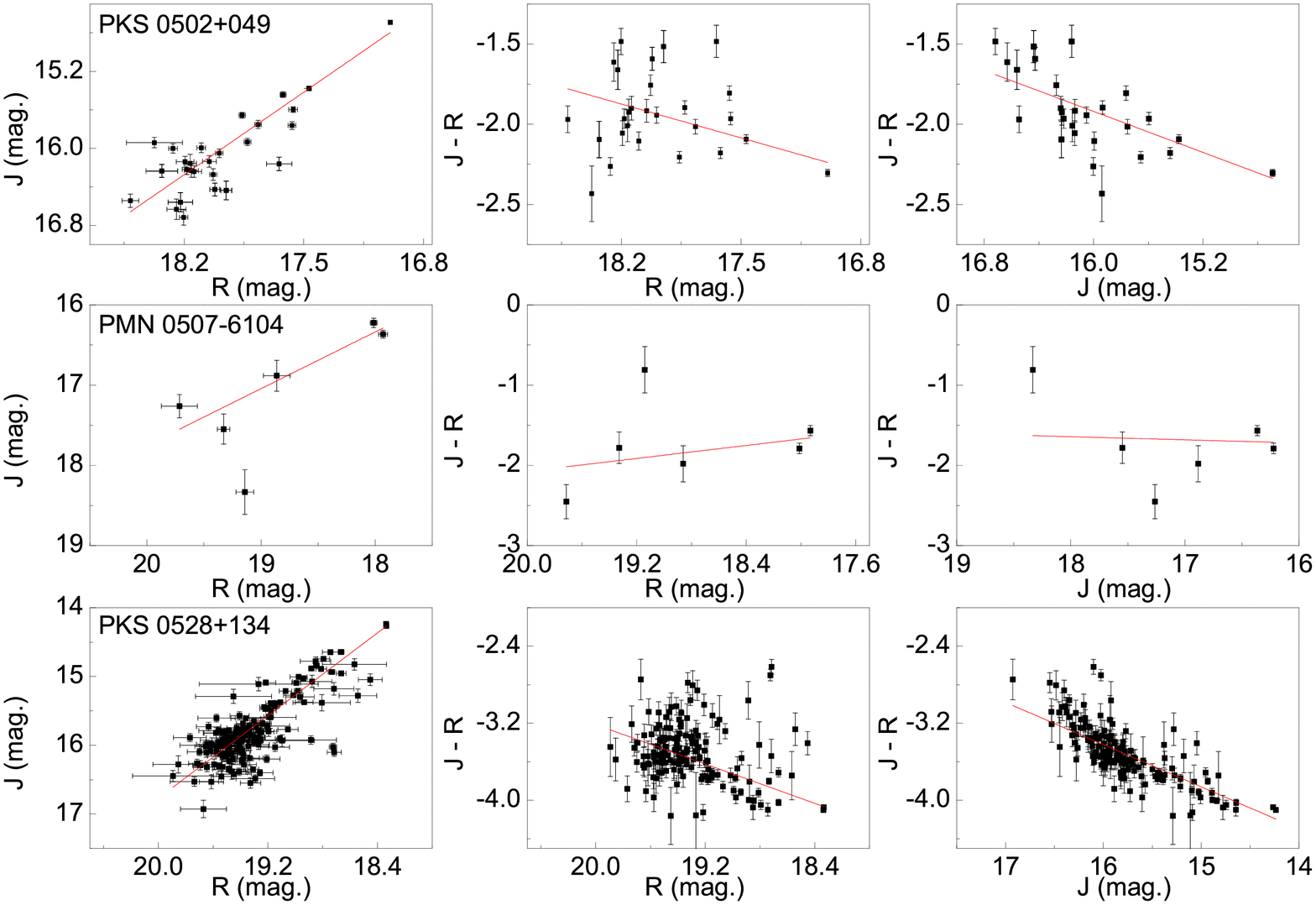}
  \includegraphics[width=6.5cm]{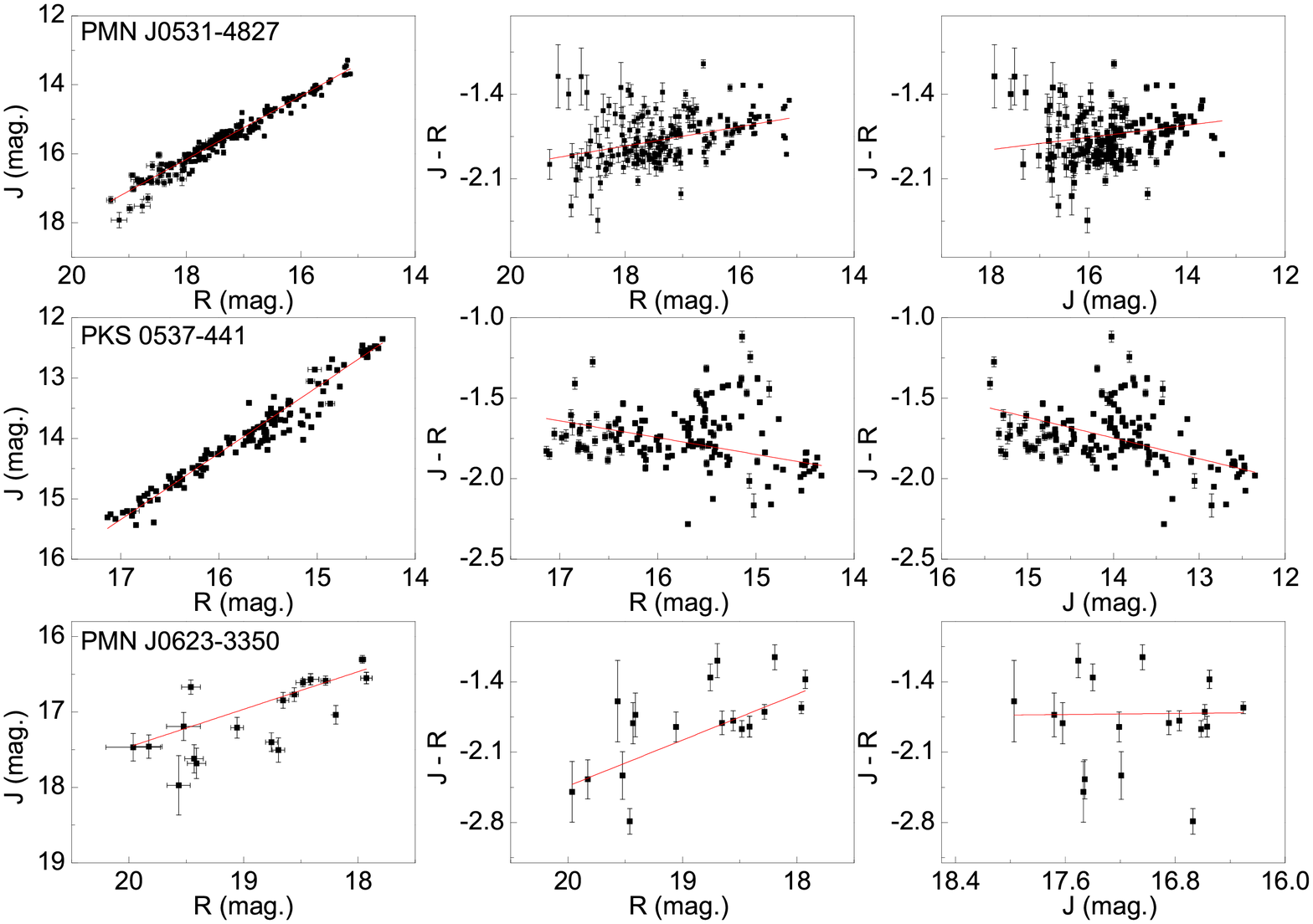}
  \includegraphics[width=6.5cm]{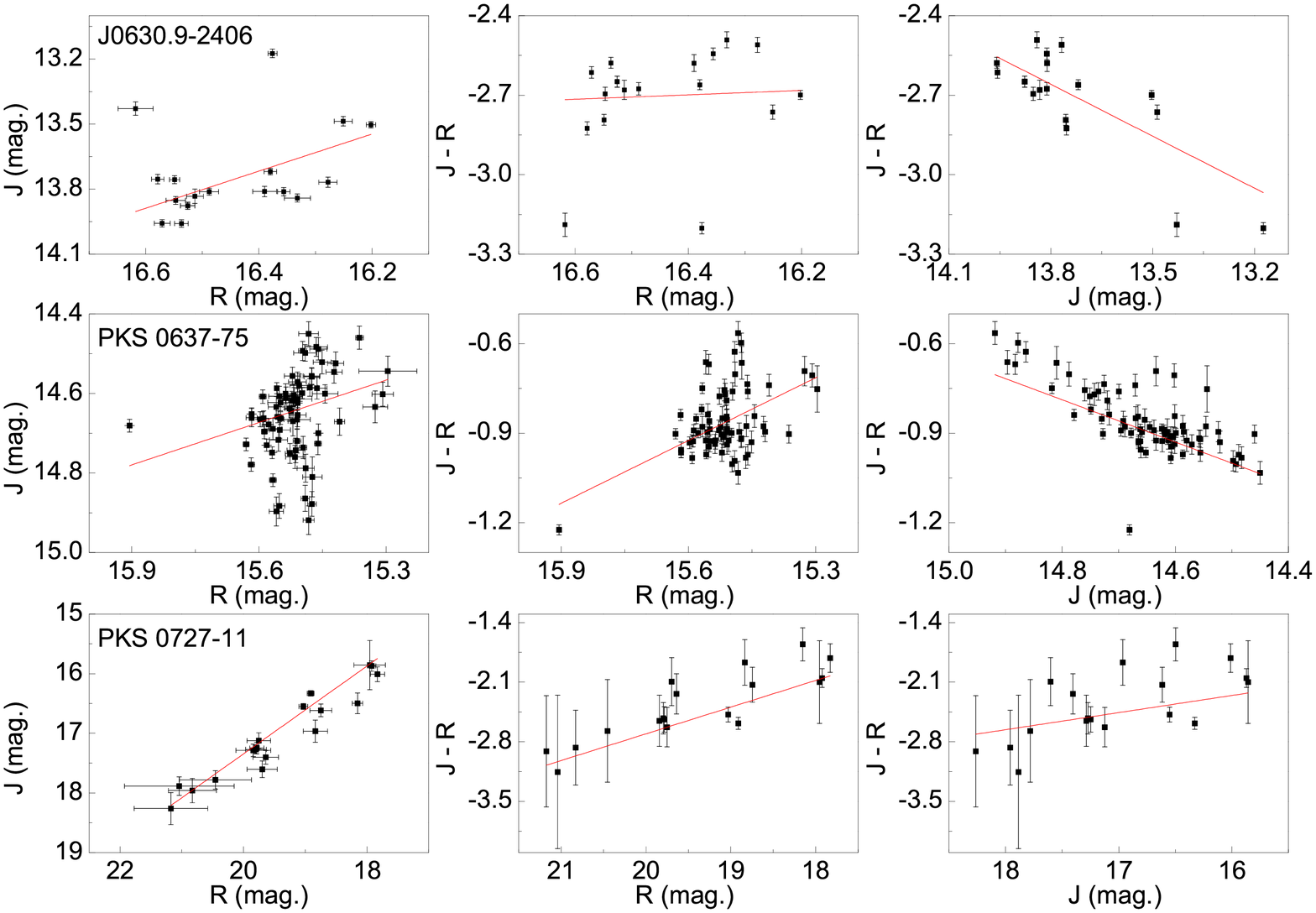}
  \includegraphics[width=6.5cm]{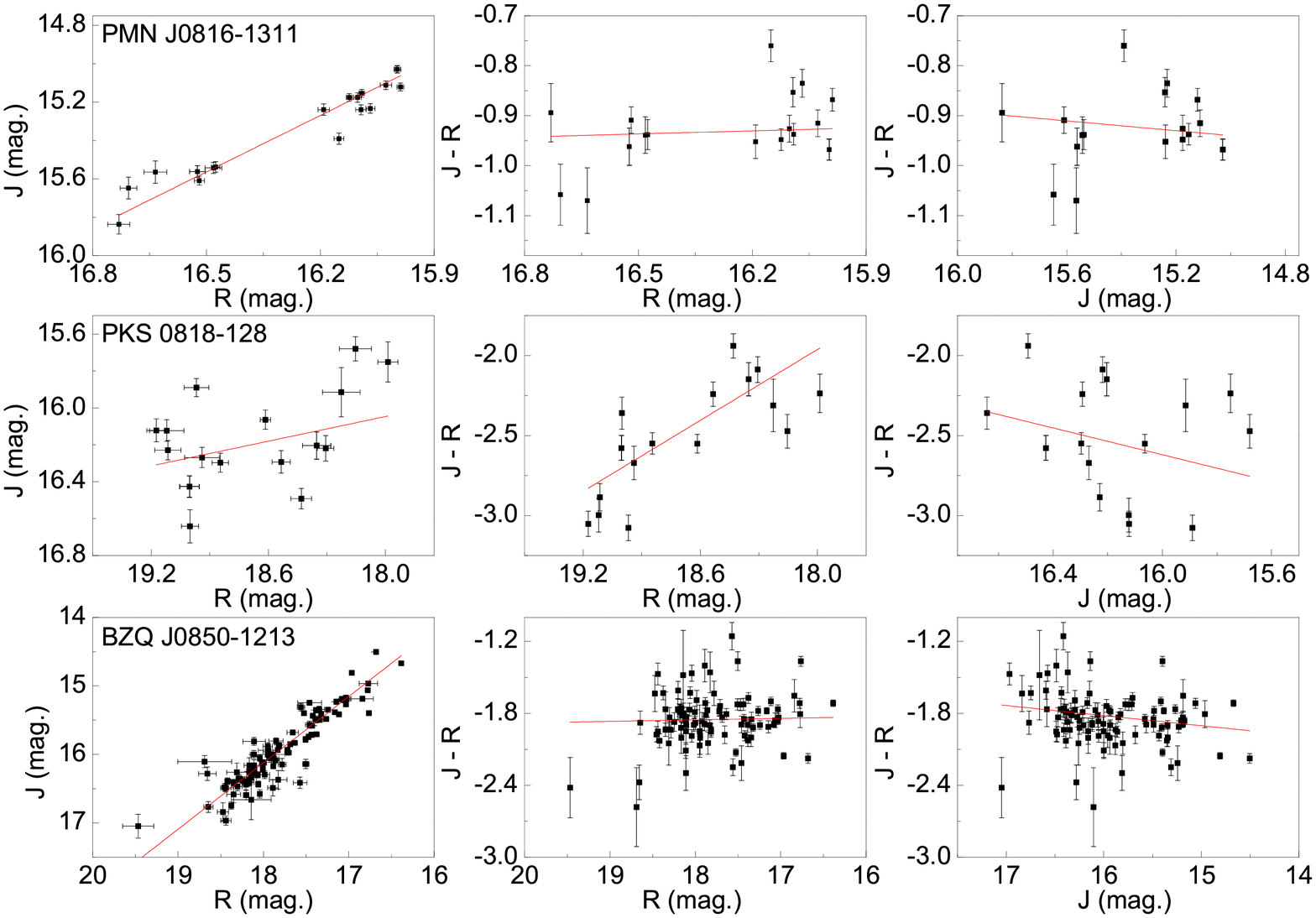}
  \includegraphics[width=6.5cm]{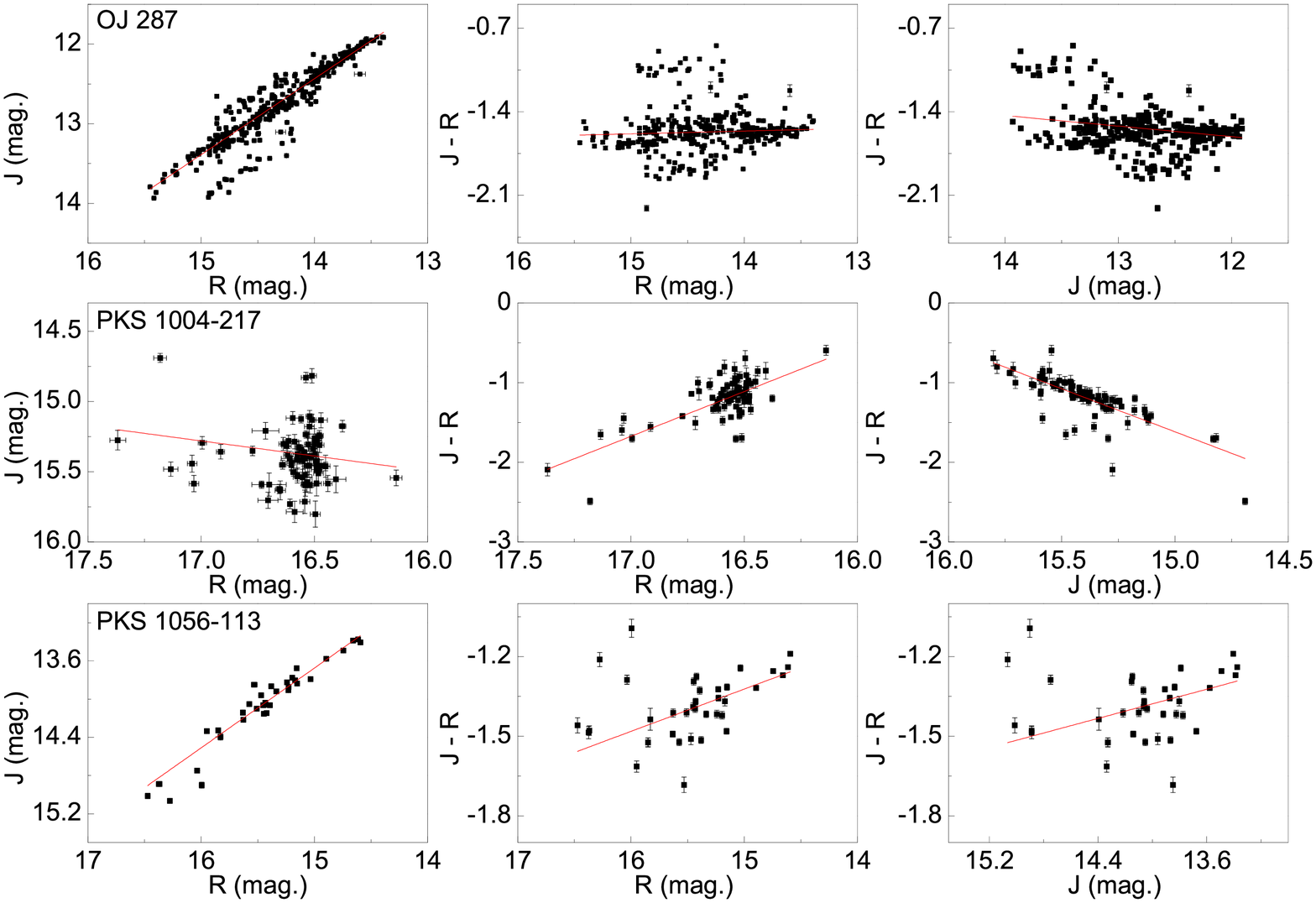}
 \caption{Continued.}
\end{figure}

\addtocounter{figure}{-1}
\begin{figure}
\centering
  \includegraphics[width=6.5cm]{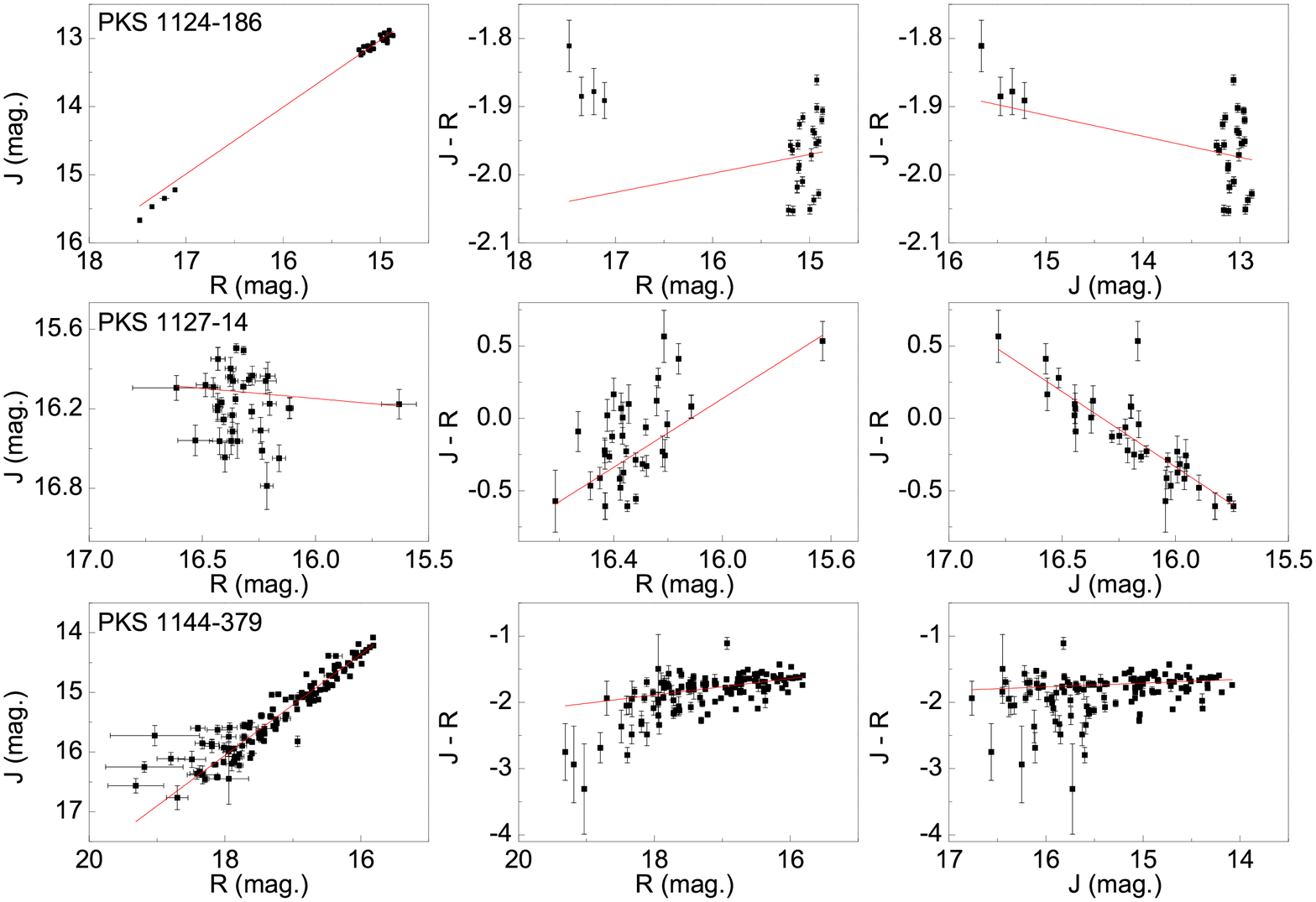}
   \includegraphics[width=6.5cm]{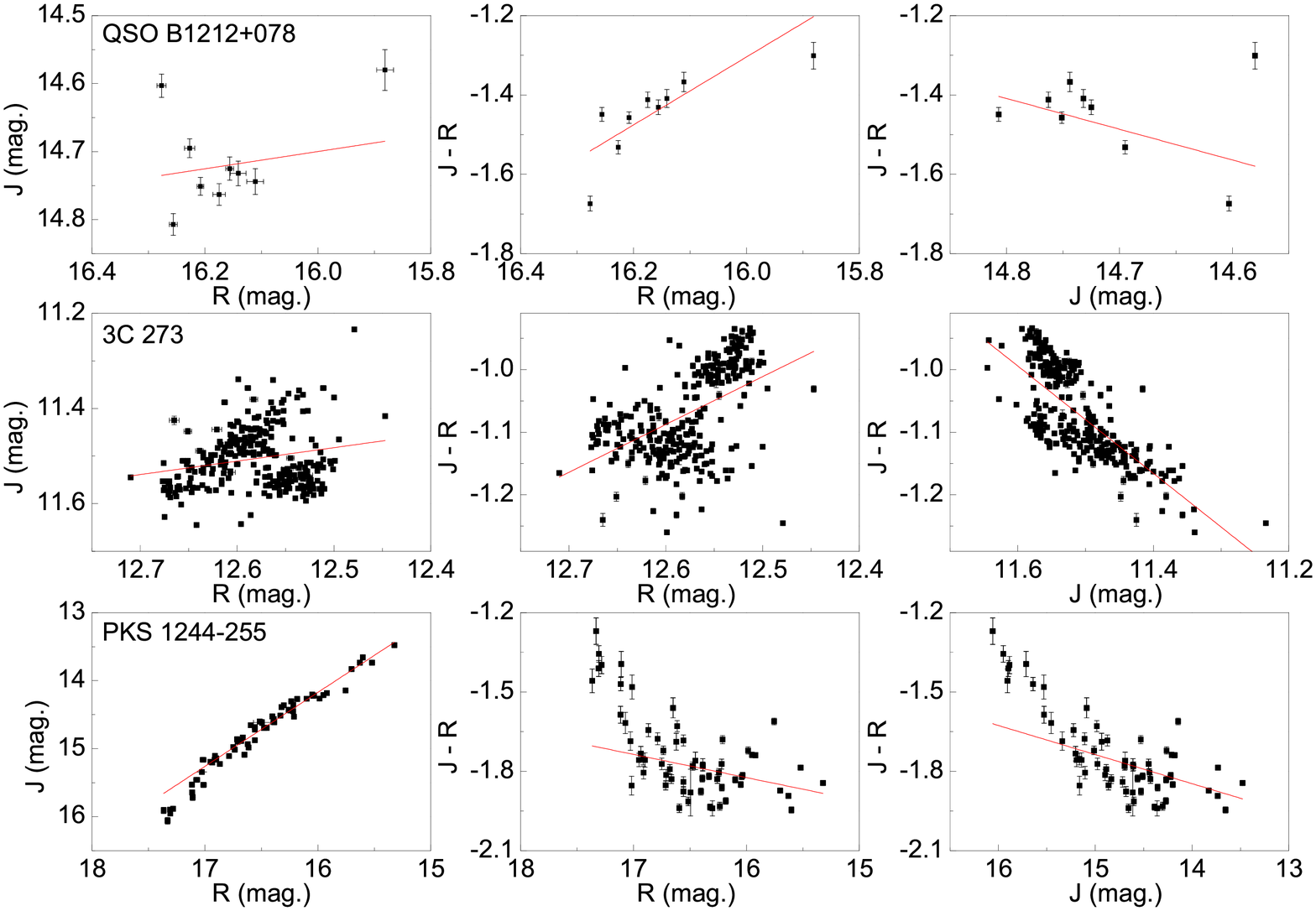}
 \includegraphics[width=6.5cm]{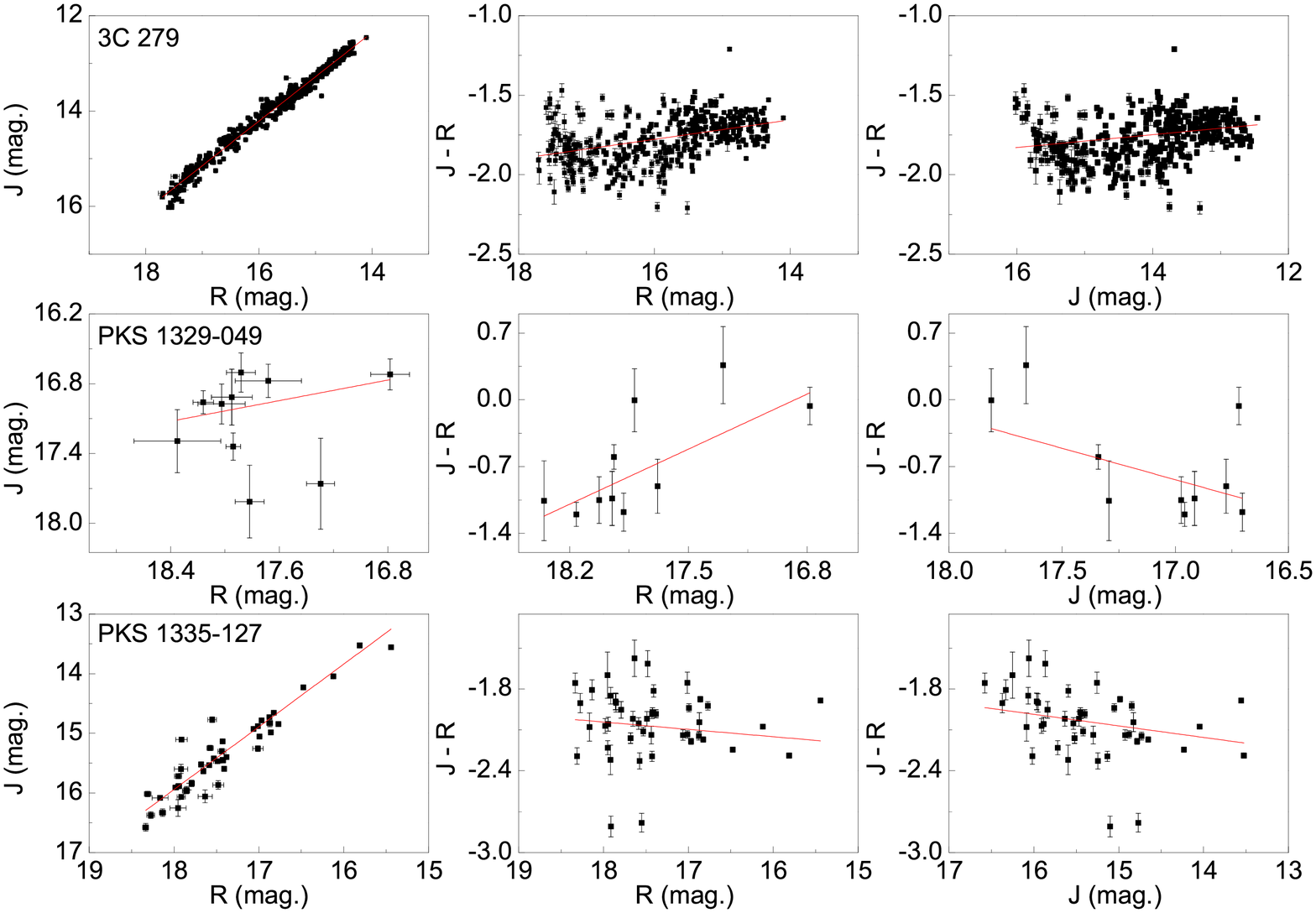}
 \includegraphics[width=6.5cm]{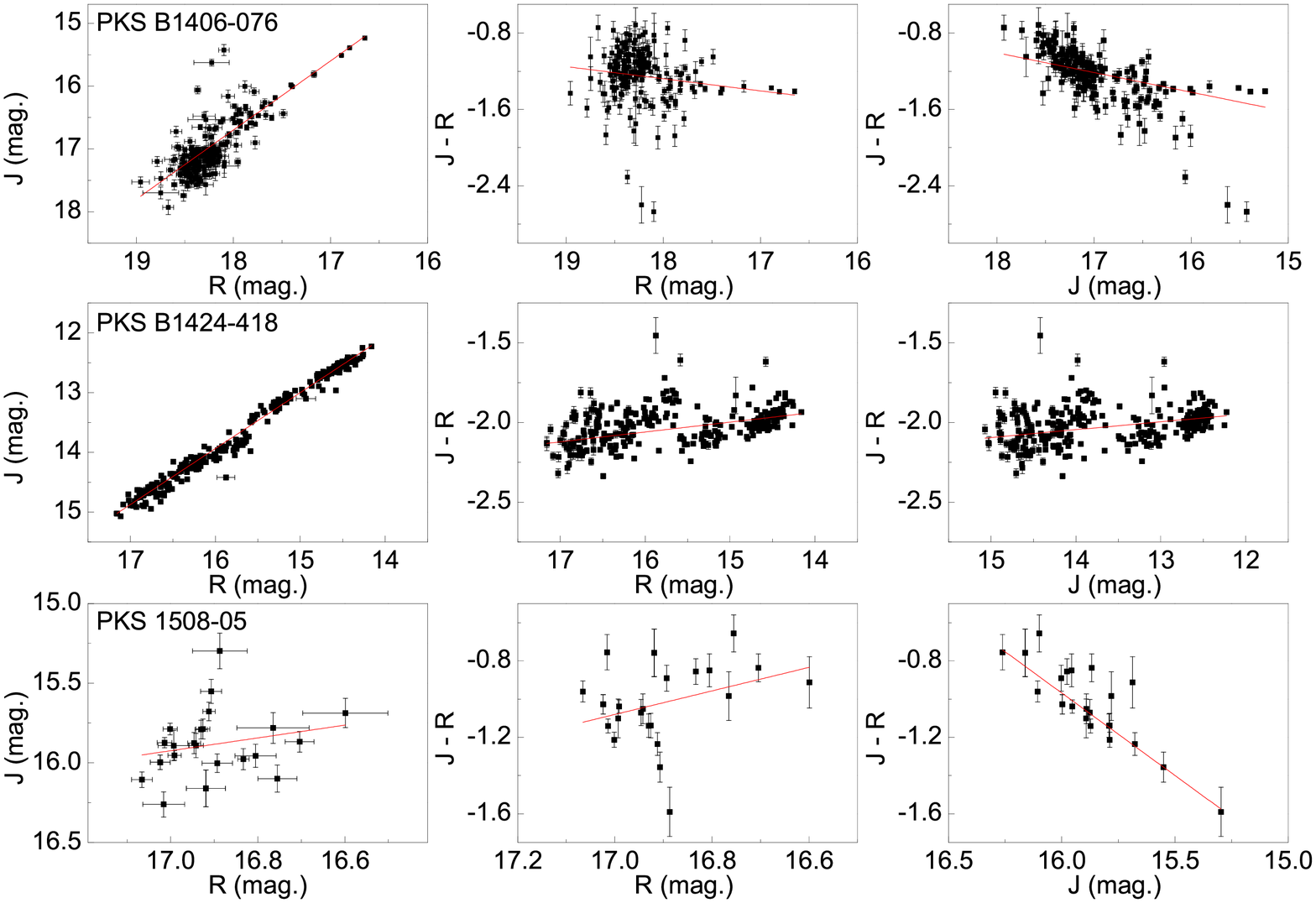}
 \includegraphics[width=6.5cm]{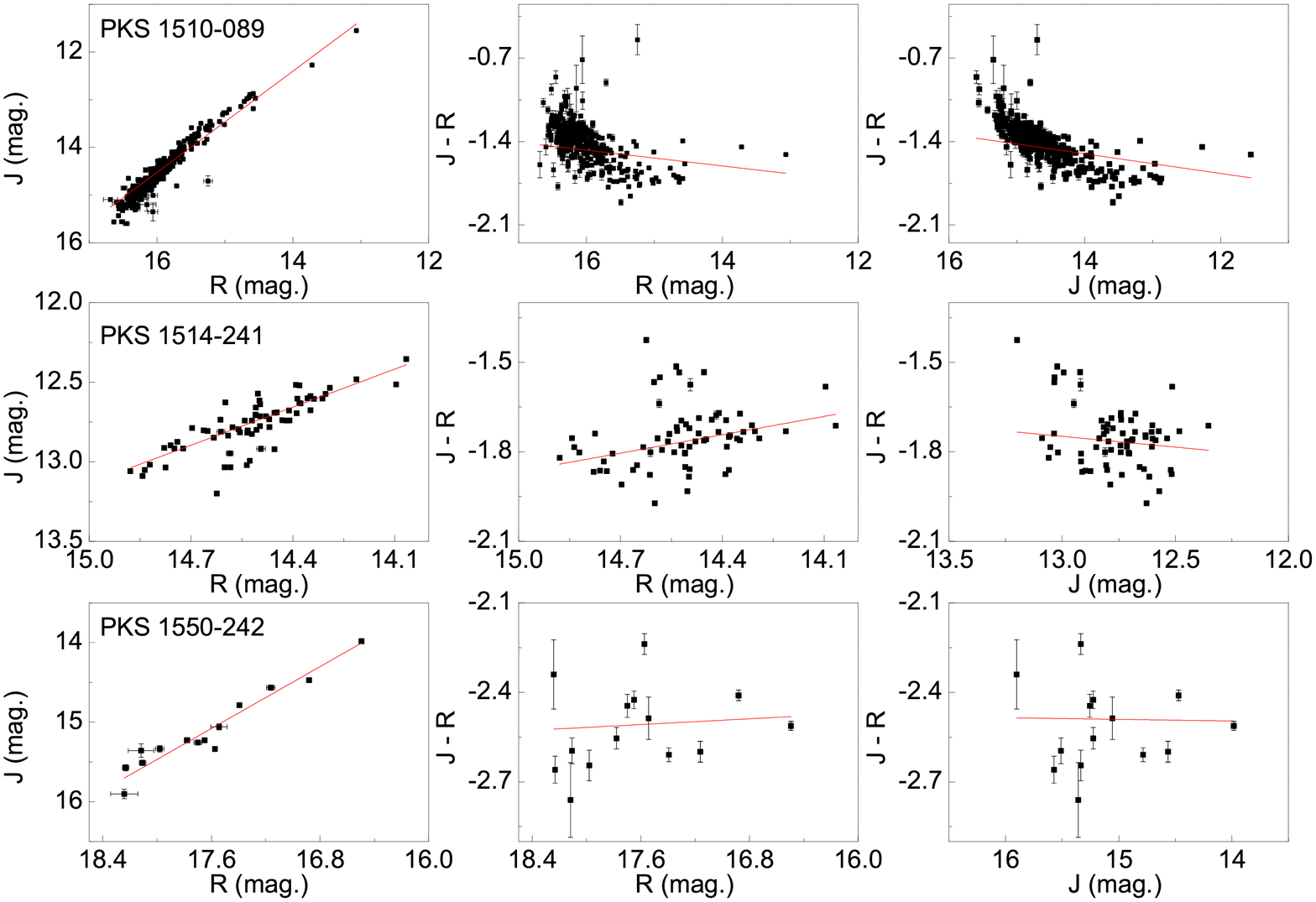}
  \includegraphics[width=6.5cm]{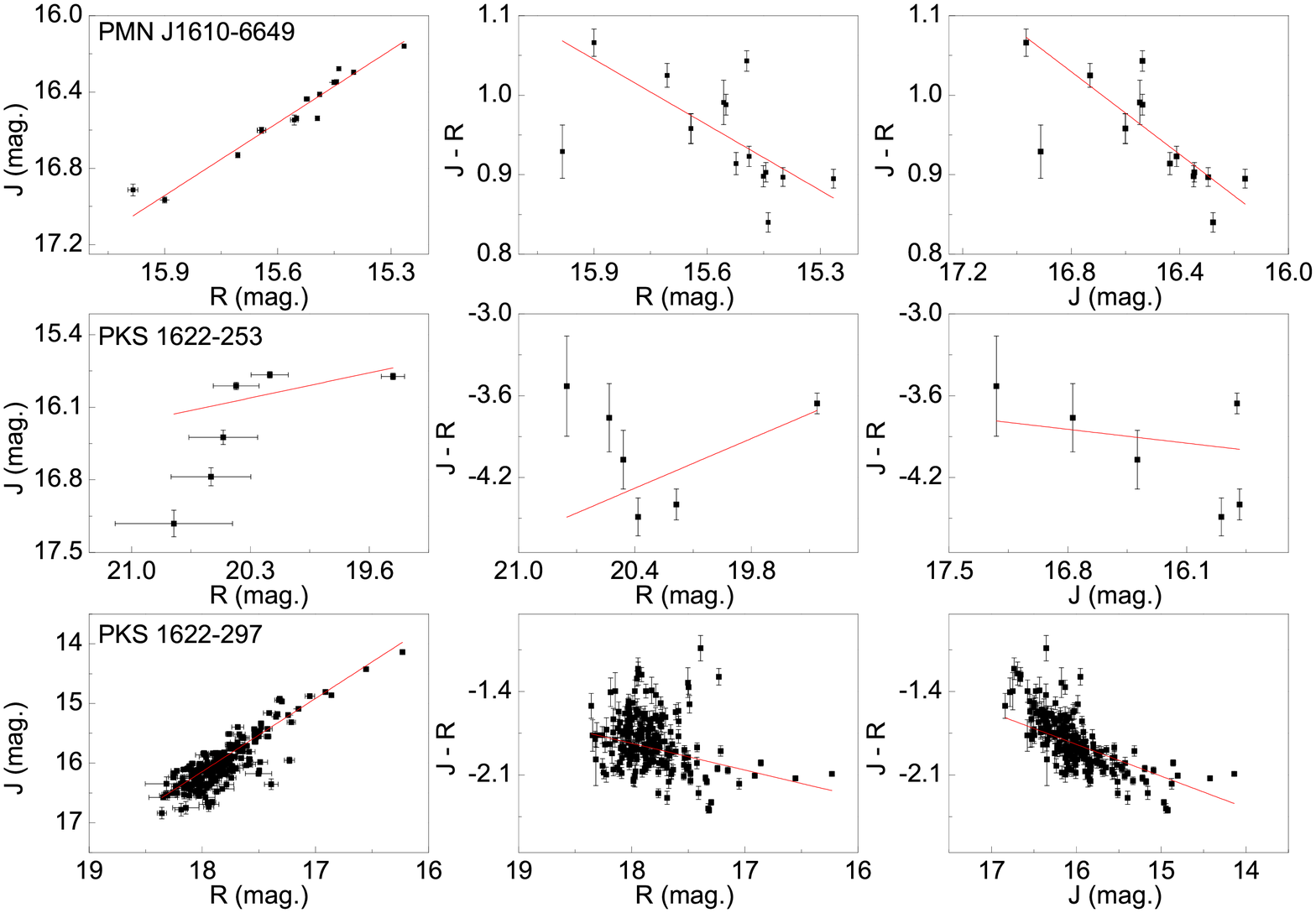}
  \includegraphics[width=6.5cm]{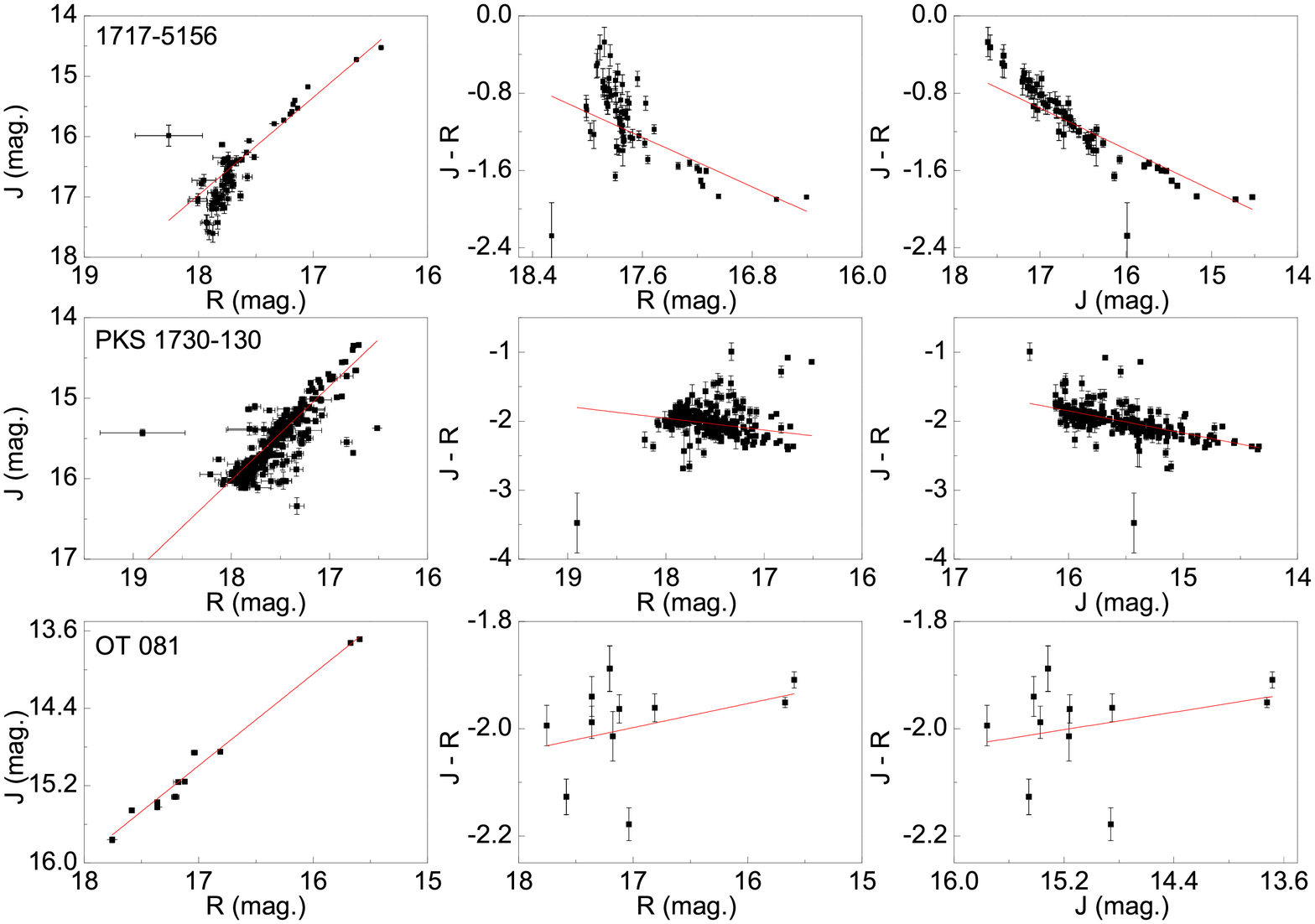}
  \includegraphics[width=6.5cm]{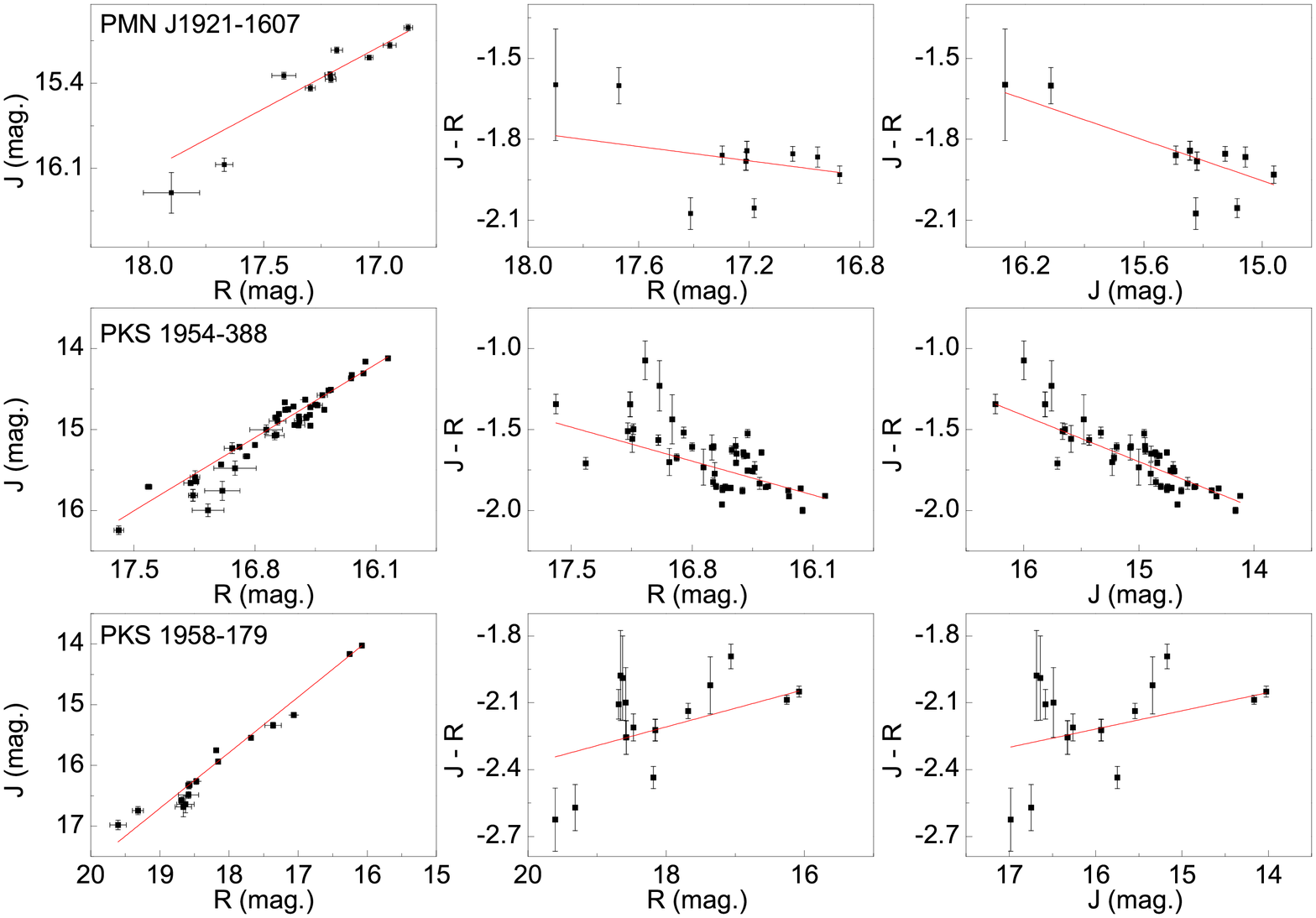}
  \includegraphics[width=6.5cm]{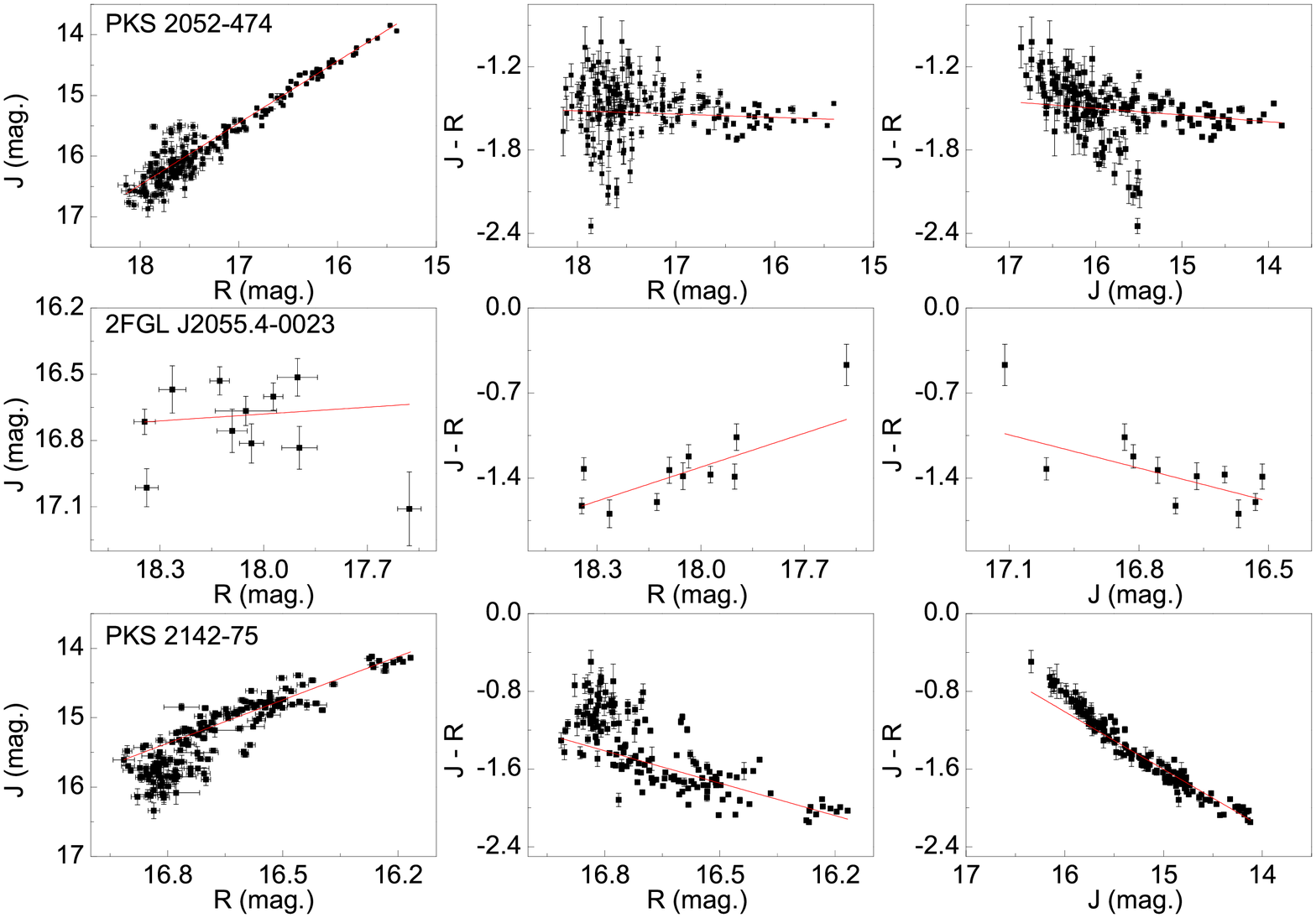}
  \includegraphics[width=6.5cm]{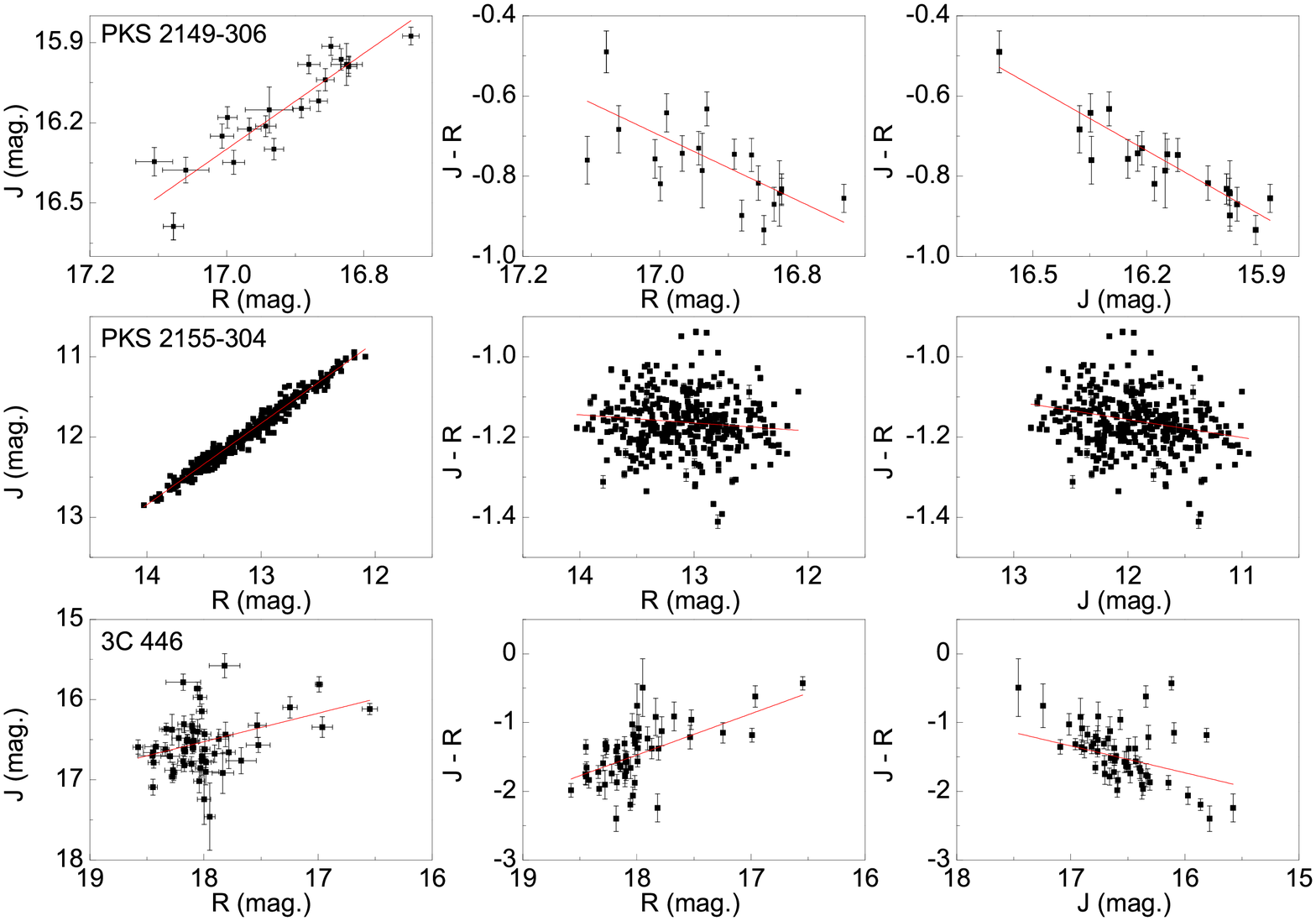}
 \caption{Continued.}
\end{figure}

\addtocounter{figure}{-1}
\begin{figure}
\centering
 \includegraphics[width=6.5cm]{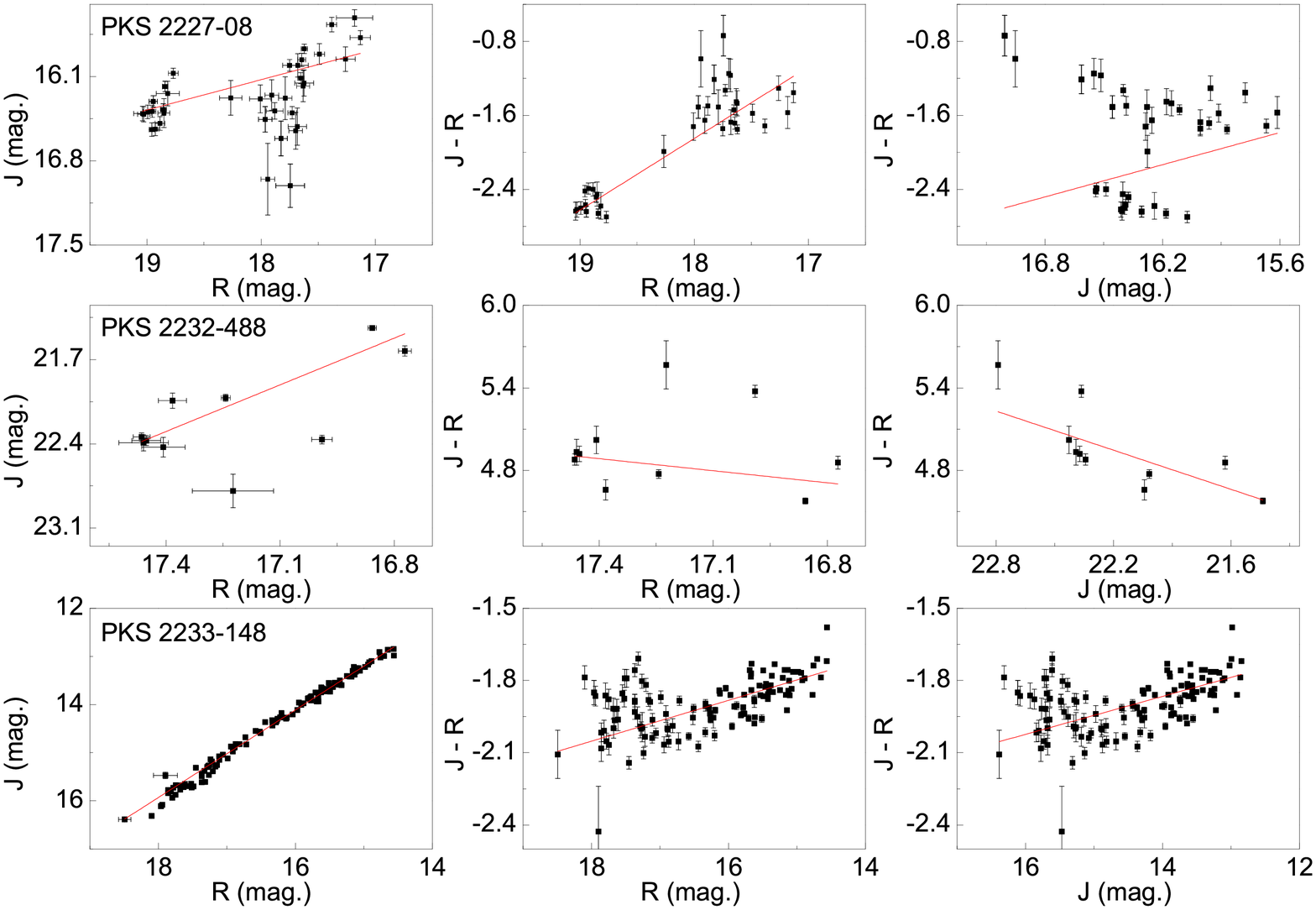}
  \includegraphics[width=6.5cm]{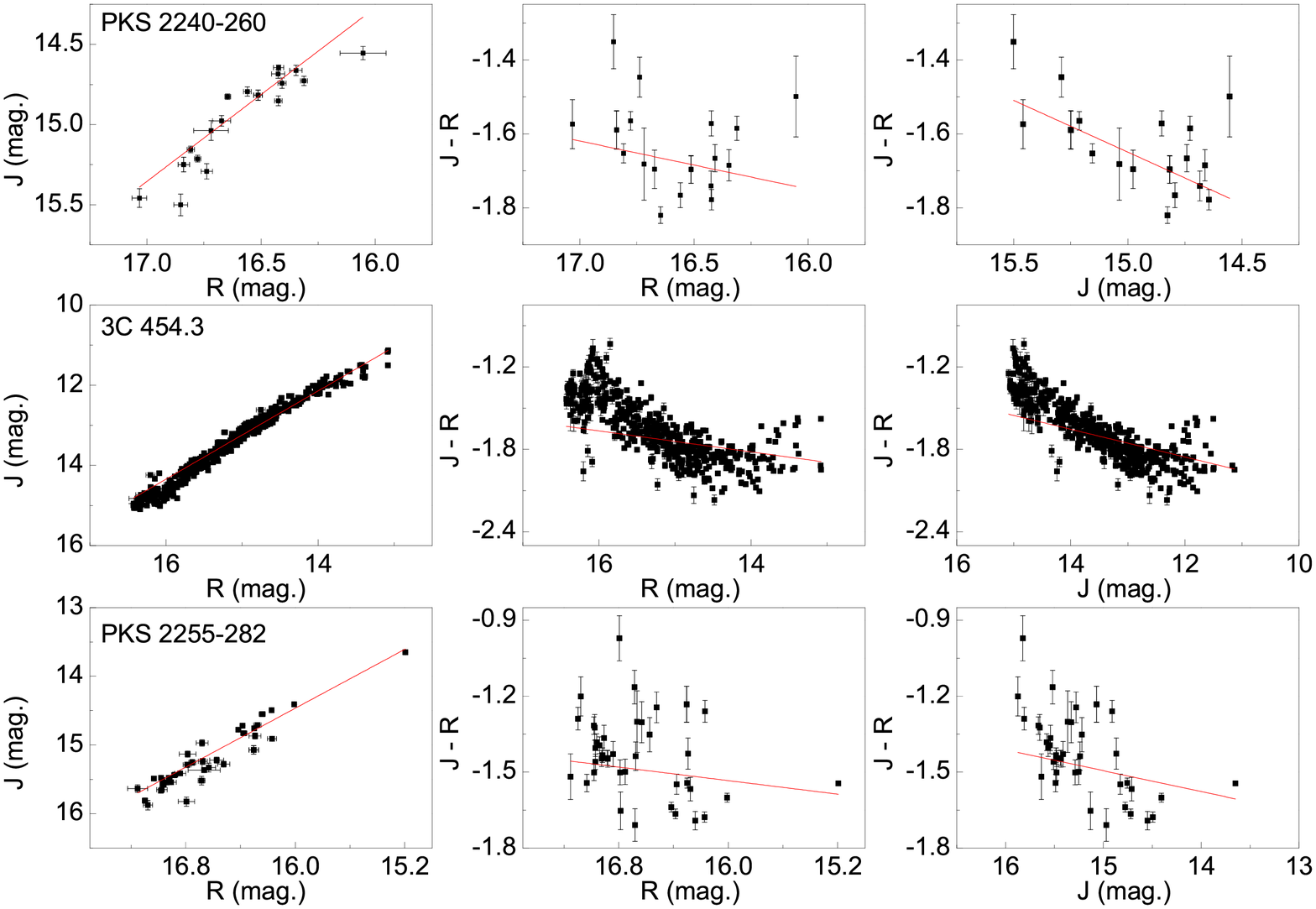}
  \includegraphics[width=6.5cm]{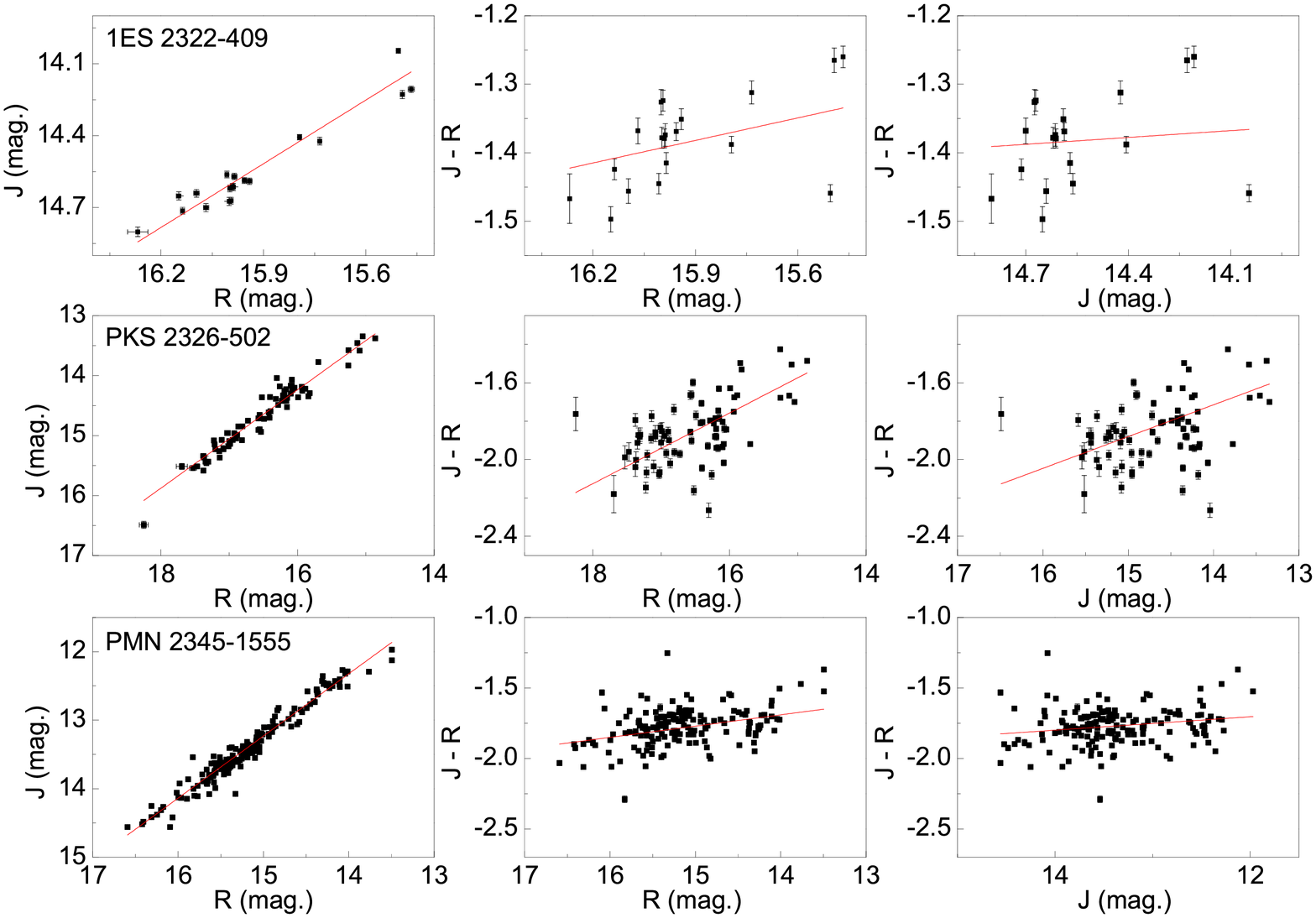}
   \includegraphics[width=6.5cm]{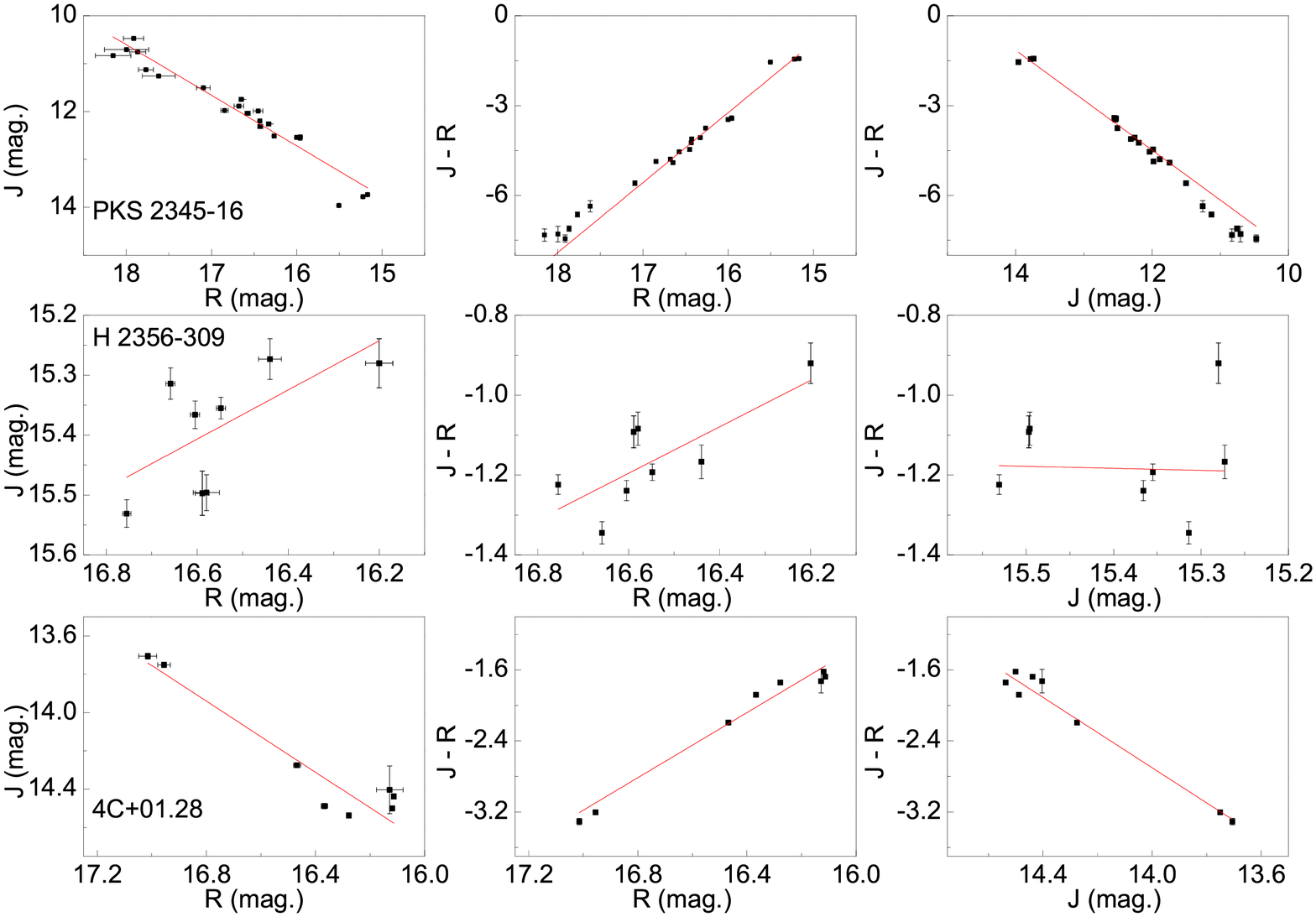}
 \caption{Continued.}
\end{figure}

\section{Discussion and conclusions}

   The BWB behavior is usually observed in BL Lac objects, which can usually be explained with shock-in-jet models. This behavior maybe means the presence of two components contributing to the overall emission, a variable component with a constant and relatively blue color and an underlying red component \citep{fioru04,ikeji11}.
RWB behavior is generally observed in FSRQs, which can be interpreted as due to a strong contribution of a blue thermal emission from the accretion disc, which mainly affects the bluer region of the optical spectrum when the jet emission is faint \citep{villata06,rani10,bonning12}.

An interesting result is that 10 sources show one color behavior in one band and an opposite trend in another band. When the $R$ band (higher frequency) brightness increases, they tend to be BWB trends. However, they show RWB behavior when the source brightness increases in the $J$ band (lower frequency). This phenomena had ever clearly appeared in the case of 3C 273 \citep{dai09}. This may be caused by the brightness variations in the higher frequency leading those in the lower frequency. In other words, the radiation emitted at higher frequencies will emerge sooner while the maximum flux at lower frequencies will be observed later \citep{villata00,papadakis04,villata09,zhang10,rani10}. Thus the variations in lower frequencies can not keep step with those in high frequencies. One would observed weak or none, even negative correlations between the two bands. Therefore, the object exhibits BWB trend in the higher frequency band, while RWB trend in the lower frequency band. We check the brightness correlations between $J$ and $R$ bands, and find that the correlations are weak or negative for these sources.

\begin{figure}[]
  \centering
   \includegraphics[width=59mm,height=58mm]{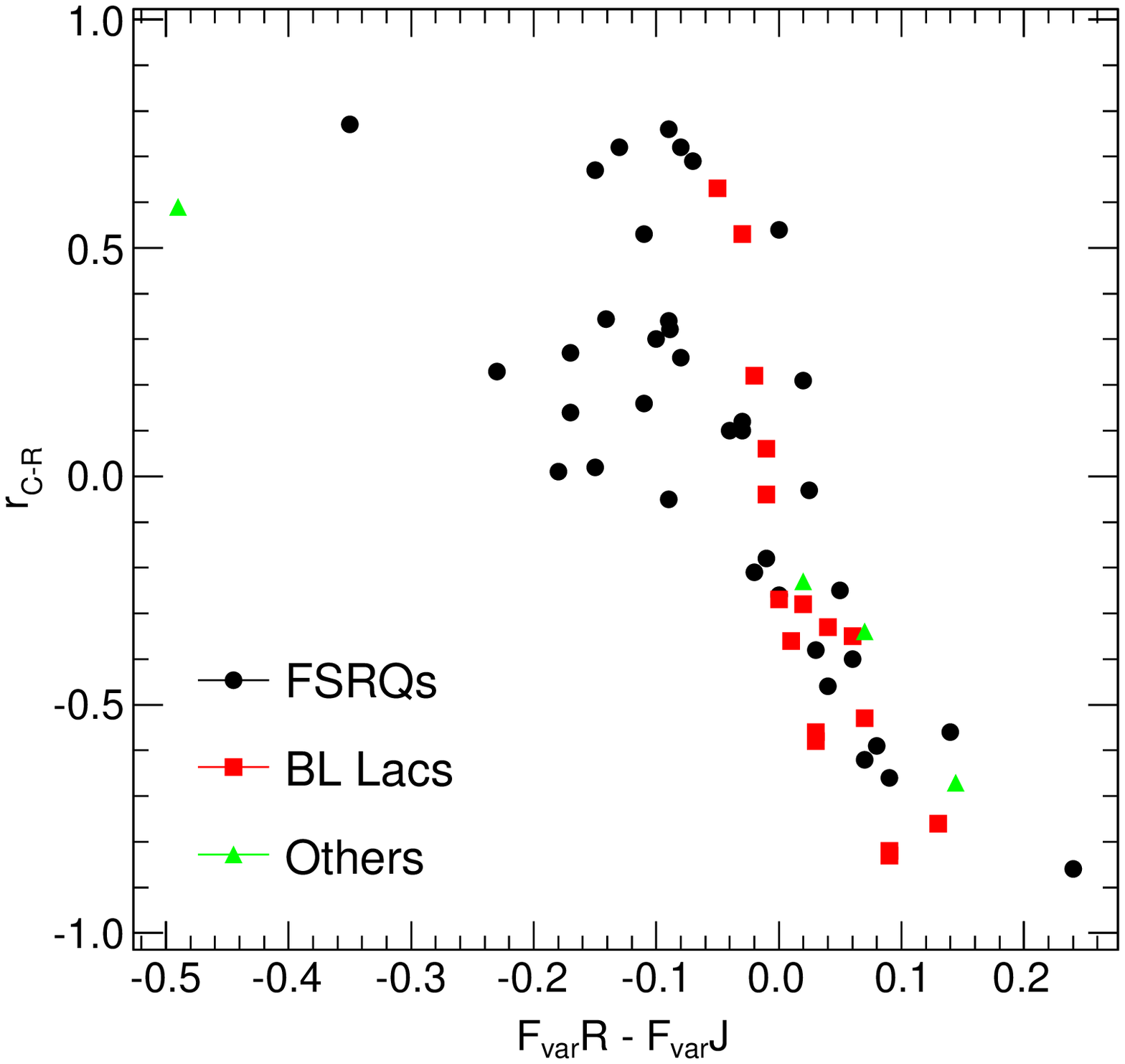}
  \caption{{\small Correlation coefficient $r_{C-R}$ versus the F$_{var}$ difference of $R$ and $J$ bands.}}
   \label{color-fvar}
\end{figure}

For the sources with strong brightness correlations between $R$ and $J$ bands, their color behaviors in the $R$ band are similar to those in the $J$ band, showing either BWB or RWB behavior.
 We investigate the relation between color behavior and $F_{var}$ difference of $R$ and $J$ bands for those objects with the absolute value of $r_{J-R}$ greater than 0.5.  In Fig.~\ref{color-fvar}, we plot the $r_{C-R}$ against the $F_{var}$ difference, $F_{varR} - F_{varJ}$. Overall, if $F_{var}$ of $R$ band is greater than that of $J$ band, $r_{C-R}$ $<0$, which means the source showing BWB trend. If $F_{var}$ of $R$ band is less than that of $J$ band, the source shows RWB trend ($r_{C-R}>0$). This means that the difference in variation amplitude can cause different color behaviors.  For an object with the fluxes at different frequencies varying simultaneously,  if the amplitude is larger at the higher frequency than at the lower one, it exhibits the BWB trend. Oppositely, the larger amplitude variations at lower frequency can result in RWB trend.

From Fig.~\ref{color}, one can see that not all sources can be fitted well by a straight line.  The color variation of some sources maybe change trend at a certain magnitude. Several sources show RWB in the low state, and then keep a  stable trend in the high state (SWB), such as PMN 0017$-$0512,  PKS 0208$-$512, PKS 1244$-$255, PKS 1510$-$089, PKS 1622$-$297 and 1717$-$5156. Some sources show RWB trend in the low state, and BWB trend in the high state, like AO 0235+164, PKS 1056-113, PKS 2233-148, PKS 2240-260 and 3C 454.3.
This behavior has also been found in other literature, such as 3C 454.3, PKS 1510-089, PG 1553+113 and PKS 0537-441 in \cite{villata06}, \cite{ikeji11} and \cite{zhang13}.
Most of these sources are FSRQs.  They have a common feature that they exhibit RWB trend in the faint state and then SWB or BWB trend in the bright state. This may be explained as the contribution reversal of different emission components. In the faint state, the strong contribution of thermal emission from an accretion disk results in RWB trend. While in the bright state, the non-thermal jet emission is dominant, which would display usual BWB trend, therefore balancing or exceeding the RWB trend. Thus, the source transforms its color behavior from RWB trend to SWB or BWB trend.

 A few sources show two or more different strips in the $Magnitude-Magnitude$ or $Color-Magnitude$ diagrams. PKS 0426-380 shows two clear different behaviors. In $J-R$ diagram, it seems to be a gap of two light curves with different traces. In the $Color-Magnitude$ diagram, there is also an intersection of two curves. One exhibits BWB trend, and another exhibits RWB trend.
OJ 287 indicates two parts in the color behavior, and the data points are very scattered. The color seems to be no trend in the $Color-R$ diagram, however, show a slight RWB trend in the $Color-J$ diagram. The BWB trend had been found in this source \citep{vagne03,fioru04,wu06,dai11}. A tendency to move around a circle in $Color-Magnitude$ diagram was also found by \cite{bonning12} and \cite{sandrinelli14}.
For the 3C 273, $J-R$ plot shows two separative parts, and the $Color-Magnitude$ plot also shows two separative parts. Two parts both have a RWB trend, and totally, $Color-J$ show RWB trend. However, in $Color-R$ plot, the combination of two part shows BWB trend. Very recently, \cite{fan14} also found that this source exhibited two separative parts in the plot of spectral index versus flux density.
PKS 2052-474 seems to display two different behaviors in the variations both with RWB trends. PKS 2227-08 shows two separative parts, which both show RWB trend. However, the combination of two parts shows a BWB trend in $C-R$ plot. These color behaviors are puzzling, which suggest that there maybe exist more complicated and separative emission components in these sources.

In conclusion, the color variability has been investigated with a large sample of blazars in optical and infrared region.
On the whole, 46 out of 49 FSRQs show color variations with RWB trends (35 FSRQs) or BWB trends (11 FSRQs), and 18 out of 22 BL Lacs show RWB trends (7 BL Lacs) or BWB trends (11 BL Lacs). Among them, 10 blazars follow RWB trend in the faint state and then SWB or BWB trend in the bright state.

\normalem
\begin{acknowledgements}
We express our thanks to the referee for great helps.
This work has been supported by the National Natural Science Foundation of
China (Grant No. 11273008), and partly by the
Foundation for Young Talents in College of Anhui Province (Grant No.
2012SQRL116) . This paper has made use of up-to-date SMARTS
 optical/near-infrared light curves that are available
 at www.astro.yale.edu/smarts/glast/home.php.

\end{acknowledgements}

\bibliographystyle{raa}

\end{document}